\documentclass{iopjournal}
\renewcommand{\articletype}[1]{}
\usepackage{amsmath}
\usepackage{amsfonts}
\usepackage{amssymb}
\usepackage{ragged2e}
\usepackage{soul}
\usepackage{braket}
\usepackage{bm}
\usepackage{upgreek}
\usepackage{mathtools}
\usepackage{bbm}
\usepackage{hyperref}

\justifying
\begin{document}
\thispagestyle{plain}

\title{Zigzag Graphene Nanoribbons as a Majorana Platform}

\author{Ruize Ma$^1$, Michele Pizzochero$^{2,3,\dagger}$, and Gaurav Chaudhary$^{4,5,*}$}

\affil{$^1$Department of Physics, ETH Z\"urich, Zurich 8093, Switzerland}

\affil{$^2$Department of Physics, University of Bath, Bath BA2 7AY, United Kingdom}

\affil{$^3$School of Engineering and Applied Sciences, Harvard University, Cambridge, MA 02138, United States}

\affil{$^4$TCM Group, Cavendish Laboratory, University of Cambridge, Cambridge CB3 0US, United Kingdom}

\affil{$^5$School of Physics, University of Bristol, Tyndall Avenue, Bristol BS8 1TL, United Kingdom}

\affil{$^{\dagger,*}$ Authors to whom any correspondence should be addressed.}

\email{$^{\dagger}$\href{mailto:mp2834@bath.ac.uk}{mp2834@bath.ac.uk}, $^*$\href{mailto:gaurav.chaudhary@bristol.ac.uk}{gaurav.chaudhary@bristol.ac.uk}}


\begin{abstract}
Graphene nanoribbons support a variety of electronic phases that can be tuned via external stimuli. In particular, zigzag graphene nanoribbons exhibit an antiferromagnetic insulating ground state which can transition to a half-metallic phase under a transverse electric field or when embedded in hexagonal boron nitride.
Here, we develop a simple model of a heterostructure composed of a half-metallic zigzag graphene nanoribbon coupled with an Ising superconductor. We demonstrate that an Ising superconductor's $s$-wave spin-singlet pairing gap component can induce a spin-triplet, odd-parity pairing gap in the half-metallic phase of the nanoribbon via proximity coupling.
The resulting superconducting phase is topologically nontrivial, with gate-tunable transitions that enable the emergence of Majorana zero modes.
\end{abstract}

\section{Introduction}
Given their potential in realizing fault-tolerant topological quantum computing, the pursuit of Majorana excitations has generated interest across condensed matter and quantum technology~\cite{Nayak2008,Kitaev2001,Kitaev2003,Alicea2012}. 
A time reversal symmetry ($ \mathcal{T}$)-broken superconductor with odd-parity pairing order parameter in its topologically non-trivial phase can host isolated Majorana excitations as topologically protected boundary states~\cite{Read2000,Ivanov2001}. 
Because naturally occurring $\mathcal{T}$-broken superconductors are rare, much effort has been devoted to the identification of heterostructures that leverage superconducting proximity effects in order to engineer topological superconductivity in an otherwise normal material. 
Notable examples include topological insulator surface states~\cite{Fu2008,Cook2011}, semiconductor quantum wires~\cite{Lutchyn2010,Oreg2010,Potter2011}, ferromagnetic atom chains~\cite{Nadj2013,Nadj2014}, and quantum (anomalous) Hall  systems~\cite{SanJose2015,Zeng2018}. 
Due to the fermionic antisymmetry of the Cooper pair wavefunction, superconductivity in a single species of fermion must exhibit an odd-parity order parameter, creating suitable conditions for $\mathcal{T}$-broken topological superconductivity with isolated Majorana zero modes (MZMs). 
In light of this, inducing superconductivity in half-metals is a promising route to realize MZMs~\cite{Chung2011,Duckheim2011}.
\begin{figure}[t]
  \centering
  \includegraphics[width=0.75\textwidth]{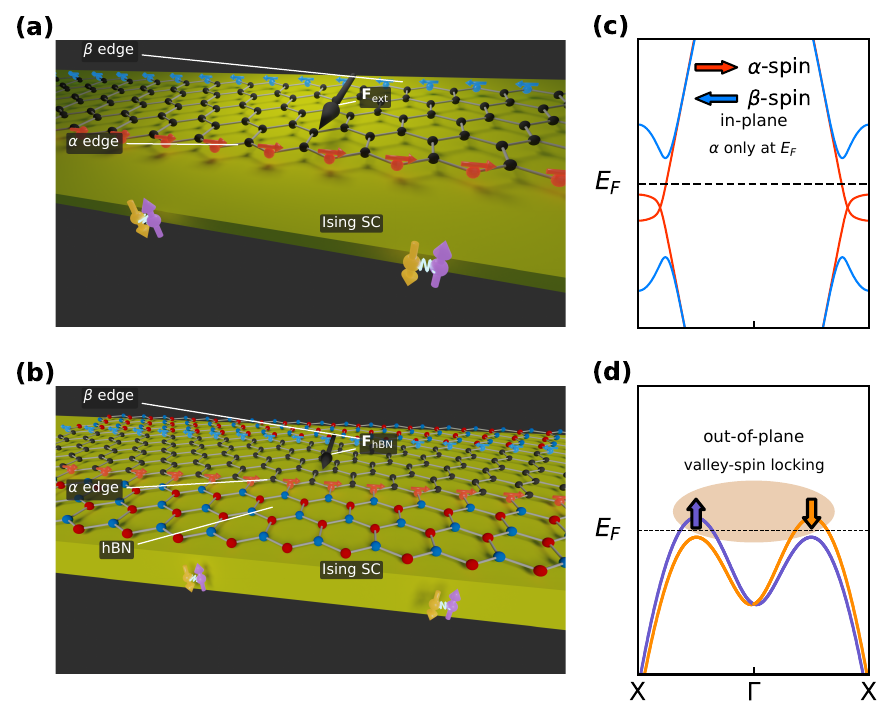}
  \caption{\label{Fig.:1_schematic}
  Schematic of the setup and the band structures of the systems considered. Zigzag graphene nanoribbons host edge antiferromagnetism, where the two opposite edges host states with spin pointing in opposite directions (as shown by red and blue arrows in (a) and (b).
  (a) A Zigzag graphene nanoribbon (ZGNR) on superconductor subjected to a transverse in-plane electric field. (b) A ZGNR embedded in hexagonal boron nitride (hBN). 
  The electric field or charge transfer from (to) the surrounding hBN breaks the symmetry between the two edges, lifting one of the spin channels to higher energy. 
  Here $\vec{F}_{\text{hBN}}$ is the induced electric field due to the hBN embedding. 
  (c) The band structure of ZGNR embedded in hBN, with the Fermi level away from neutrality. The key feature is the half metallic bands with $\alpha$ spin. 
  (d) The qualitative bands of the underlying metallic phase of the Ising superconductor. The key feature is the spin locking with the band.
  We have chosen a notation such that $\alpha$ and $\beta$ are spin states polarized with a quantization axis in the plane of the heterostructure and $\uparrow, \downarrow$ are spin states with polarization with an out-of-plane quantization axis.}
\end{figure}

In this letter, we propose a highly tunable Majorana platform based on half-metallic zigzag graphene nanoribbons (ZGNRs)~\cite{Son2006,Tepliakov2023}. ZGNRs are quasi-one dimensional strips of graphene that have been recently fabricated in atomically precise fashion through on-surface synthesis \cite{Ruffieux2016}. Owing to a Stoner-like instability resulting from a van Hove singularity at the Fermi level, they exhibit $\pi$-electron magnetism. Local magnetic moments are localized at the edges and couple antiparallel across the axis of the nanoribbon, giving rise to a spin-0 antiferromagnetic ground state \cite{Yazyev2010}. The signature of this magnetic order has been experimentally detected~\cite{Blackwell2021,Brede2023}, even up to room temperature~\cite{Magda2014}. Transverse electric fields -- either external \cite{Son2006}, as shown in Fig.\ \ref{Fig.:1_schematic}(a), or built-in fields \cite{Tepliakov2023}, e.g., upon embedding ZGNRs into hexagonal boron nitride \cite{Chen2017, Wang2021b}, as shown in Fig.\ \ref{Fig.:1_schematic}(b) -- induce an inversion-symmetry breaking which lifts the degeneracy of the edge states of the nanoribbon, enforcing a half-semimetallic band structure at charge neutrality that can be readily gated to the half-metallic phase depicted in Fig.\ \ref{Fig.:1_schematic}(c). However, the suitability of such half-metallic nanoribbons to realize MZMs, and the extent to which they can support them, has so far remained unexplored.

Here, we consider half-metallic ZGNRs proximity coupled to an Ising superconductor, as schematically illustrated in Fig.\ \ref{Fig.:1_schematic}(a,b). As a model material systems, we consider ``Ising superconductors", such as monolayer $\text{NbSe}_2$ and $\text{MoS}_2$,  where the spin-orbit (SO) coupling locks the spin to valley with an out-of-plane anisotropy, resulting in an anomalously large in-plane critical field~\cite{Lu2015,Xi2016}. We show that owing to the broken inversion symmetry, the the spin-singlet gap component of the Ising superconductor can induce triplet odd parity pairing gap in the half-metallic state of
the ZGNR, mapping the system to a Kitaev chain ~\cite{Kitaev2001}, without the assistance of explicit non spin-conserving
terms. By tuning the gate voltage or external electric field, transitions between between trivial and topological phases can be achieved, thus acting as a knob to create and manipulate MZMs.

\section{Proximity-induced superconductivity in half-metals} 
In half-metals, the Fermi level crosses the electronic bands of one spin orientation, while residing within the energy gap of the opposite spin orientation~\footnote{Given recent discoveries of half and quarter metal phases in rhombohedral trilayer graphene \cite{Zhou2021}, we emphasize that for the case considered, there is no extra degree of freedom such as valley. Hence, the half-metal nomenclature refers to a single spin-polarized band intersecting the Fermi level}, as shown in Fig.\ \ref{Fig.:1_schematic}(c). 
Naively, one does not expect a heterojunction comprising a spin-singlet $s$-wave superconductor and a half-metal  to support proximity-induced superconductivity. 
Because the half-metal admits only one spin orientation at the Fermi level, only the electron with the same spin orientation forming the Cooper pair can tunnel into it, with the opposite-spin electron being reflected. Yet, we demonstrate that this picture breaks down under specific conditions. In the following, we derive the minimal conditions under which the superconductivity can still be induced in the half-metal upon proximity coupling to a singlet $s$-wave superconductor. For simplicity, we consider a one-dimensional Hamiltonian of the heterostructure where the half-metal and superconductor are coupled via a single fermion tunneling,
\begin{align}\label{Eq:HamHJ1}
    H = H_{HM} + H_{SC} + H_T,
\end{align}
where 
\begin{subequations}\label{Eq:HamHJ2}
\begin{align}
    & H_{HM} = \sum_k \epsilon(k) \hat{\chi}^{\dagger}_{k,s} \hat{\chi}_{k,s} , \label{Eq:HamHM}\\
    & H_{SC} = \sum_{k,\sigma} \xi(k) \hat{\gamma}^{\dagger}_{k,\sigma} \hat{\gamma}_{k,\sigma}  + \sum_{k} \Delta ( \hat{\gamma}^{\dagger}_{k,\uparrow} \hat{\gamma}^{\dagger}_{-k,\downarrow} - \hat{\gamma}^{\dagger}_{k,\downarrow} \hat{\gamma}^{\dagger}_{-k,\uparrow}) + \text{h.c.} , \label{Eq:HamSC} \\
    & H_T = \sum_{k,\sigma} T_{\sigma,s}(k) \hat{\chi}^{\dagger}_{k,s} \hat{\gamma}_{k,\sigma} + \text{h.c.} \label{Eq:HamTunnel}
\end{align}
\end{subequations}
In these expressions, $\hat{\chi}^{\dagger}_{k,s}$ creates a spin `$s$' electron in the half metal, $\hat{\gamma}^{\dagger}_{k,\sigma}$ creates a spin `$\sigma$' electron in the superconductor, $\Delta$ is an even-parity spin-singlet pairing gap in the superconductor, $T_{\sigma,s}(k)$ is the band projected single electron tunneling between the superconductor and the half metal that can have a non-universal form and complicated $k$-dependence depending on the microscopic details of the system, and $\epsilon(k)$ [$\xi(k)$] is the energy dispersion of the half-metal (normal phase of the superconductor) evaluated from the Fermi level. 
To the second-order in perturbation, we can integrate out the tunneling Hamiltonian $H_T$ and obtain an effective Hamiltonian in the half-metal region
\begin{align}\label{Eq:HamHMSC}
    \tilde{H}_{HM} = \sum_k \tilde{\epsilon}(k) \hat{\chi}^{\dagger}_{k,s} \hat{\chi}_{k,s} + \biggl( \tilde{\Delta}(k) \hat{\chi}^{\dagger}_{k,s}\hat{\chi}^{\dagger}_{-k,s} + \text{h.c.} \biggr ),
\end{align} 
where $\tilde{\epsilon}(k)$ is renormalised dispersion and 
\begin{align}\label{Eq:GapInduced}
    \tilde{\Delta}(k) =  \frac{2\Delta[T_{\uparrow,s}(k) T_{\downarrow,s}(-k) - T_{\downarrow,s}(k) T_{\uparrow,s}(-k)]}{4\Delta^2 - [\epsilon(k) - \xi(k)]^2}
\end{align}
is the proximity-induced gap (see Supplementary Note 1 in ~\cite{SM}). 
The divergence in this expression is not physical, indicating the limitation of the second-order perturbation approximation.
The induced gap in the half metal must be odd-parity, as required from the fermionic antisymmetry of the Cooper pair, which is also satisfied by the RHS of Eq.~\ref{Eq:GapInduced}. However, if the spin $s$ of the half-metal and the spin $\sigma$ of the superconductor have the same quantization axis, then spin-flip tunneling processes are required to induce non-vanishing pairing. 
Moreover, even with the spin-flip tunneling, inversion symmetry further leads to a vanishing proximity gap in Eq.~\ref{Eq:GapInduced}, as the two terms in the numerator cancel each other under inversion symmetry. 
Thus, establishing broken inversion symmetry as a necessary criterion to induce pairing in the half metal from a nearby singlet $s$-wave superconductor.
Fortunately, broken inversion symmetry at the interfaces can lead to Rashba SO coupling, which in turn allows for spin-flip tunneling effects, since spin is no longer a conserved quantum number~\cite{Gorkov2001,Edelstein2003}. Indeed, prior studies have considered interfacial Rashba and SO coupled superconductor, respectively, to induce a pairing gap in the half-metal \cite{Chung2011, Lee2009, Duckheim2011}.

We discuss another route to achieve a finite proximity gap. Both $\uparrow$ and $\downarrow$ spins from the superconductor can tunnel to the half-metal, provided that its spin quantization axis is not collinear with the superconductor. For example, if $\sigma$ is quantized transverse to the heterojunction plane and $s$ is polarized in the plane of the heterojunction, then $T_{\uparrow,\rightarrow}/T_{\uparrow,\uparrow} = - T_{\downarrow,\leftarrow}/T_{\downarrow,\downarrow} = 1/\sqrt{2}$. 
Broken inversion symmetry is nevertheless still required for a finite induced gap. 
To this end, Eqs.~\ref{Eq:HamHJ1} and \ref{Eq:HamHJ2}, only specify the generic low energy expansion near the respective Fermi levels of the two subsystems, i.e. capturing the half metallicity and spin-singlet even-parity superconductivity. 
Because of the need for broken inversion symmetry, from now on, we consider the superconducting subsystem to be that of an Ising type. The Ising SO coupling (details in Supplementary Note 2 \cite{SM}) is intrinsic to many two-dimensional transition metal dichalcogenides with non-centrosymmetric crystals~\cite{Xiao2012}. 
Similar to the Rashba SO coupling, in the Ising SO coupling the $\text{SU(2)}$ spin rotation symmetry is broken. Unlike the Rashba SO coupling, however, at the Fermi level, spins are collinear and remain a conserved quantum number, as illustrated in Fig.~\ref{Fig.:1_schematic}(d)\footnote{The Ising SO coupling renders spin non-colinearities required to connect the two valleys of opposite spin at higher energies away from Fermi level}. 

Contrary to earlier studies on half metals \cite{Chung2011, Duckheim2011}, in our work we propose the realization of an induced pairing gap in a half-metal proximitized to an Ising superconductor containing only spin-singlet $s$-wave pairing gap, without relying on spin-flip tunneling or other spin conservation breaking terms. Since intrinsic Ising spin-orbit couplings are typically stronger than Rashba or spin-flip scattering terms, we expect such induced pairing gaps to be larger than those in, e.g., Refs.~\cite{Chung2011, Duckheim2011}, possibly offering a practical advantage for future device applications. 
Previous works considering Ising superconductors to engineer MZMs in half-metals are similar in spirit to this work~\cite{Zhou2016, Lesser2020}. However, they attributed the proximity induced pairing to intrinsic triplet pair correlations of the Ising superconductor~\cite{Zhou2016, Lesser2020} . Although such correlations are expected to be present in Ising superconductors, there is no evidence that the actual superconducting phase has triplet pairing gap component. In contrast, here we attribute the proximity induced gap to arise from the singlet $s$-wave gap component of the Ising superconductor, which still induces pairing gap in the half metal due to non-trivial $k$ dependence in band project inter-layer hopping matrix elements, satisfying the broken inversion symmetry requirement of Eq.~\ref{Eq:GapInduced}.

\section{ZGNR-Ising superconductor heterostructure} 
We consider a tight-binding (TB) mean-field Hubbard model for the ZGNR under an in-plane electric field transverse to the periodic direction of the nanoribbon. 
The Hamiltonian of the ZGNR includes only the unhybridized $p_z$ orbitals of the $\text{sp}^2$-bonded carbon atoms on a honeycomb lattice with nearest-neighbor hopping~\cite{Pizzochero2024, Ma2025}.
\begin{align}\label{Eq:Ham_GNR}
    & \mathcal{H}_{\text{GNR}} = -t_0\sum_{\langle i, j \rangle, s} (\hat{c}^{\dagger}_{i,s}\hat{c}_{j,s} + \text{h.c.})  + \sum_{i,s} (\epsilon_i + U \langle \hat{n}_{i,-s} \rangle - \mu) \hat{c}^{\dagger}_{i,s}\hat{c}_{i,s},
\end{align}
where $\hat{c}^{\dagger}_{i,s}$ creates a $p_z$ electron with spin $s$ at the ZGNR site $i$. The first term describes the kinetic energy with the nearest neighbor hopping $t_0=2.88\ \text{eV}$. The second term describes on-site energy that has contributions from: (i) the site-dependent $\epsilon_i = F \cdot y_i$ originating from the transverse electric field of strength $F$, where $y_i$ is the transverse coordinate, (ii) the global chemical potential $\mu$, and (iii) mean-field interaction potential, where $U = 2.88\ \text{eV}$ is the on-site Hubbard interaction, and $\langle \hat{n}_{i,s} \rangle$ is the self-consistently determined mean-field spin density. The $U/t \approx 1$ is a standard choice within the range estimated for $\pi$-conjugated carbon systems from \textit{ab initio} calculations and correctly captures the emergence of edge magnetism in ZGNRs~\cite{Pisani2007}, which also approaches the experimentally inferred values obtained for $\text{sp}^2$-carbon chains~\cite{Thomann1985}. 
In Supplementary Note 5 \cite{SM}, we examine the dependence of our findings on $U/t$. For narrow nanoribbons under moderate electric fields, our predictions indicate that the MZMs are expected for interaction strengths of $U/t \gtrapprox 0.7$, pointing to their robustness to modest screening effects in the most experimentally relevant regimes.
In Supplementary Note 6 \cite{SM} we consider the sensitivity of our results to spatially varying interlayer tunneling, which serves to emulate the effect of lattice mismatch in the heterostructure. We still find significant regions with MZMs, suggesting that the lattice mismatch does not significantly impact their stability.

We describe the minimal electronic structure of the Ising superconductor through a spinful two-orbital Hamiltonian 
\begin{align}\label{Eq:IsingMetal}
    & \mathcal{H}_{\text{SC}} = \sum_{i,j}\sum_{\tau,\tau',\sigma} t^{\tau\tau'}_{ij} \hat{f}^{\dagger}_{i,\tau,\sigma} \hat{f}_{j,\tau',\sigma} + i\lambda' \sum_{i,\sigma,\tau} \sigma \tau \hat{f}^{\dagger}_{i,\tau,\sigma}\hat{f}_{i,-\tau,\sigma} \notag\\
    &\hspace{2.5cm} -\mu' \sum_{i,\tau,\sigma} \hat{f}^{\dagger}_{i,\tau,\sigma} \hat{f}_{i,\tau,\sigma} + \sum_{i,\sigma,\tau} \sigma\Delta_{\tau} \hat{f}^{\dagger}_{i,\tau,\sigma} \hat{f}^{\dagger}_{i,\tau,-\sigma} + \text{h.c.},
\end{align}
where $\hat{f}^{\dagger}_{i,\tau,\sigma}$ creates an electron at site $i$ with spin $\sigma$ in orbital $\tau$. The matrix element $t^{\tau\tau'}_{ij}$ contains the hopping amplitudes between orbital $\tau$ at site $i$ and orbital $\tau'$ at site $j$, $\lambda'$ is the SO coupling, $\mu'$ is the chemical potential, and $\Delta_{\tau}$ is a phenomenological on-site, intra-orbital pairing gap that leads to singlet $s$-wave superconductivity. 
Note that this mean-field model does not contain any attributes of triplet pair correlation that are typically present in Ising superconductor. 
Hence, any induced gap from this model necessarily originates from the singlet gap component. 
This model is grounded on the tight-binding Hamiltonian of monolayer transition metal dichalcogenides $\text{MX}_2$, where M is a transition metal and X is a chalcogen atom~\cite{Liu2013}. 
In a top-down view, monolayer $\text{MX}_2$ forms a hexagonal lattice consisting of two inter-penetrating triangular sublattices, one originating from the chalcogen atoms and the other from the metal atoms. Spin-orbit coupling, which originates from the $d_{xy}$ and $d_{x^2-y^2}$ orbitals of the metal atom \cite{Liu2013}, is the feature of interest to our purposes. Therefore, we consider a restricted model by only retaining these two orbitals in the triangular sublattice of the metal atoms. We further assume that the nearest-neighbor distance in $\text{MX}_2$ and ZGNR is identical and the two systems are perfectly aligned, with the metal atoms underneath the B sublattice of the ZGNR. Subsequently, the $\text{MX}_2$ and the ZGNR are only coupled via a vertical hopping process involving the M sites of the former and the B sublattice of the latter.
\begin{align}\label{Eq:Ham_tunneling}
    \mathcal{H}_T = \sum_{i,\tau,s,\sigma} w^{\sigma,s}_{\tau} \hat{c}^{\dagger}_{is} \hat{f}_{i\tau\sigma} + \text{h.c.} 
\end{align}

As discussed in the previous section and crucial to our setup, the spin quantization axes for $s$ and $\sigma$ are not collinear. In our calculations, the spin quantization axis of the superconductor is out-of-plane such that $\sigma = \uparrow\downarrow$, while the spin quantization axis of the ZGNR is in-plane and we refer to them as $\alpha$ and $\beta$ spins, as is customary in the graphene literature. However, we mention that usually $\alpha, \beta$ nomenclature does not specify the quantization axis. Here, we associate this terminology specifically to in-plane polarization for a clear distinction with  the spin quantization axis of the superconductor.

\begin{figure}[t]
  \centering
  \includegraphics[width=0.75\textwidth]{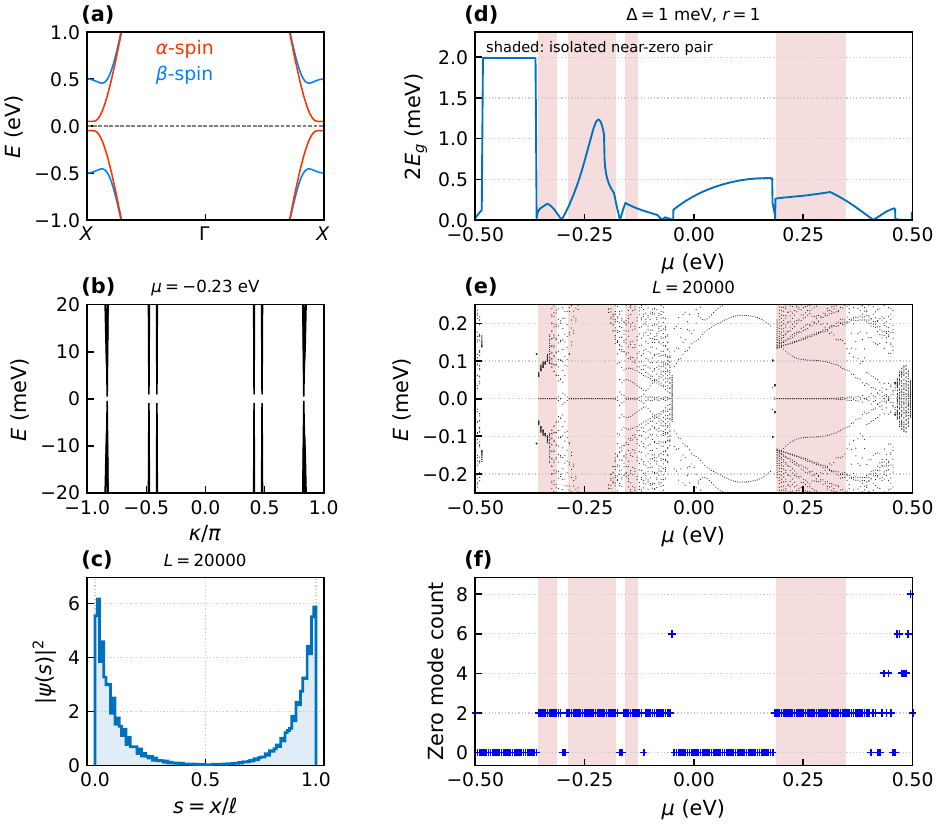}
  \caption{\label{Fig.:2_modelCalc}
  Calculations for the ZGNR-Ising superconductor heterostructure. The calculations are performed for ZGNR of width $W=3$ using the following parameters: GNR onsite Coulomb repulsion and nearest-neighbour hopping integrals $U=t=2.88\ \text{eV}$, superconducting gap $\Delta_{\tau} = 1$~meV, tunneling amplitude $|w_{\tau}| = 0.1$~eV, SC chemical potential $\mu' = 1.0$~eV, Ising spin-orbit coupling $\lambda' = 0.2$~eV, and the hopping amplitudes $t^{\tau\tau'}_{ij}$ as in Ref.~\cite{Liu2013}. (a) Bandstructure of the ZGNR in transverse in-plane electric field $F = 0.12\ \text{V/Å}$ showing magnetic bands where $\alpha$-spin bands are split from $\beta$ spin bands. (b) BdG spectrum of the heterostructure obtained at $\mu = -0.23\ \text{eV}$. 
  (c) The real space distribution $|\psi(x)|^2$ of the zero energy mode showing localization at the two ends of the nanoribbon. 
  (d) Bogoliubov excitation gap for the heterostructure with periodic boundary condition as a function of chemical potential $\mu$ shows spectral gaps and topological phase transitions indicated by gap closings and openings.
  (e) Low-energy spectrum near zero energy for a finite-length system (L=20000 unit cells). The gapped continuum of bulk states accurately mirrors the PBC gap shown in (d). Crucially, two in-gap states emerge, pinned at zero energy in the topological phase. These are the Majorana zero modes (MZMs) localized at the system's ends.
  (f) Number of zero-energy states, determined by counting eigenvalues with an energy $|E| < 10^{-5}$ eV. The plot confirms that the system hosts either zero or exactly two zero-energy modes, consistent with the expected topological invariant for this phase.
  }
\end{figure}

In our calculations, we first determine $\langle \hat{n}_{i\sigma} \rangle$ by self-consistently solving for the isolated ZGNR under an external field, following Eq.~\ref{Eq:Ham_GNR}. 
The resulting ZGNR ground state is insulating with a small gap where the nearest conduction and valence bands are spin-polarized, as in Fig.~\ref{Fig.:2_modelCalc}(a). 
Under gate control, this leads to a half-metallic ground state. 
The obtained spin densities are then treated as a fixed potential for the subsequent step where the superconductor is introduced. The full Bogoliubov-de-Gennes (BdG) Hamiltonian is thus diagonalized in a single step, without further self-consistency.

\section{Emergence of tunable topological phases} 
In Fig.~\ref{Fig.:2_modelCalc}(a), we show the band structure of the ZGNR in transverse in-plane electric field. At charge neutrality, it consists of a pair of $\alpha$-spin bands with a small gap in the vicinity of the $X$-point along with large gap $\beta$-spin bands. Remarkably similar results are obtained for the ZGNR embedded in hexagonal boron nitride~\cite{Tepliakov2023}, (see Supplementary Note 3 in \cite{SM}). By tuning the chemical potential into the $\alpha$ spin band, we achieve a half-metallic state.
In Fig.~\ref{Fig.:2_modelCalc}(b), we show a representative case, of the BdG spectrum of the setup, showing a finite BdG gap.
In Fig.~\ref{Fig.:2_modelCalc}(d), we show the Bogoliubov excitation gap of the proximity coupled system as a function of the chemical potential in the ZGNR.  Remarkably, we observe that the system is usually gapped, showing that the spin-singlet Ising superconductor is indeed effective to induce a proximity pairing gap in the half-metal even when the two systems are coupled by a simple elastic tunneling process, without invoking any spin-flip or interfacial Rashba effects. 
As argued in previous section, such induced gap in the half metal must be odd-parity, which is responsible for topologically non-trivial phases and MZMs as discussed here.
The spectrum of Bogoliubov excitations closest to zero energy in a representative finite system of $20000$ unit cells is given in Fig.~\ref{Fig.:2_modelCalc}(e). For the most part, when $\mu$ is within the $\alpha$ spin band, the system exhibits zero-energy modes separated from the rest of the spectrum by the bulk gap, as expected for this one-dimensional topological phase. 
In Fig.~\ref{Fig.:2_modelCalc}(f), we show the count of zero-energy modes. Apart from a few isolated incidences of more than two near-zero energy modes, in the whole region under consideration, there are either no zero modes, corresponding to the trivial phase, or exactly two zero modes (one mode at each end of the ZGNR), thus ensuring that the zero modes are well isolated. In Fig.~\ref{Fig.:2_modelCalc} (c), we show the real space distribution of probablity density of zero energy state, which clearly shows localisation at the two ends of the nanowire.

The findings shown in Fig.~\ref{Fig.:2_modelCalc} can be understood analytically by restricting to the two $\alpha$-spin bands $\epsilon_{\alpha\pm}(k)$, 
which is an even function of $k$ (see Supplementary Note 4 for a simpler model \cite{SM}). 
Starting with the Fermi level in the lower band and examining only the band that crosses the Fermi level, 
\begin{align}\label{Eq:BdG}
    H_{BdG}(k) = \begin{pmatrix}
        \epsilon_{\alpha,-}(k)-\mu & \tilde{\Delta}(k) \\
        \tilde{\Delta}^{\ast}(k) & -\epsilon_{\alpha,-}(k)+\mu
    \end{pmatrix},
\end{align}
is the band projected BdG Hamiltonian. Here $\tilde{\Delta}(k)$ is odd in $k$, as per Eq.\ \ref{Eq:GapInduced}, and $\tilde{\Delta}(\pm\pi)=0$ due to the lattice periodicity.
This BdG matrix can be antisymmetrized under the unitary Majorana transformation. The $\mathbb{Z}_2$ topological invariant can be obtained as $\nu = \text{sgn}[\text{Pf} \tilde{H}_{BdG}(0)]\text{sgn}[\text{Pf} \tilde{H}_{BdG}(\pi)]$~\cite{Kitaev2001}, where $\tilde{H}_{BdG}$ is in the Majorana basis and $\text{Pf}$ is the Pfaffian, giving 
\begin{align}\label{Eq:Z2Invariant}
    \nu = \text{sgn}[(\epsilon(0)-\mu)(\epsilon(\pi)-\mu)].
\end{align}

Depending on the value of $\mu$, we can distinguish various topological regimes.
For values of $\mu$ inside the band $\nu = -1$, the $\mathbb{Z}_2$ invariant is non-trivial, signaling the formation of MZMs at the end of a finite system. 
This is a remarkable property of the Kitaev chain, where the superconductor is always topologically non-trivial when the Fermi level resides inside the band. Owing to the even parity of $\epsilon_{\alpha}(k)$ and odd parity of $\tilde{\Delta}(k)$, Eq.~\ref{Eq:BdG} is identical to a Kitaev chain model and, therefore, exhibits the same topological properties. On the other hand, the Kitaev chain is topologically trivial for values of $\mu$ outside the band.  When $\mu$ intersects only the lower band $\epsilon_{\alpha,-}$, the system is in the topologically non-trivial phase. 
As a result, isolated MZMs form at the two ends of a finite system.  As $\mu$ approaches charge neutrality, the ZGNR spectrum becomes gapped. 
As a result superconductivity is not induced in the ZGNR. 
Our BdG calculations in this limit show a trivial gapped spectrum as expected, since the system can be viewed as a disconnected band insulator and a trivial $s$-wave superconductor.
As $\mu$ is further tuned to lie in $\epsilon_{\alpha,+}$ band the system transitions back to a topologically non-trivial phase with MZMs at the end.
When $\mu$ is tuned to be large enough on either the positive or the negative sides such that both $\alpha$ and $\beta$ bands cross the chemical potential, the system is topologically trivial, since the ZGNR can accommodate both spins of the Cooper pair in this case leading to a predominantly singlet $s$-wave induced pairing.
We should note that the induced gap is sensitive to the relative position of the chemical potential in the ZGNR band and the normal state of the superconductor.
 In our case, it is possible that in this limit, the induced gap is absent (too small) such that the overall system behaves as a metal with no significant BdG gap as seen in large $|\mu|$ limits. We observe such vanishing gap behavior beyond the scales presented in Fig.~\ref{Fig.:2_modelCalc} (d-f) for $|\mu|>0.5 \text{eV}$. We do not expect this region to be favorable for MZMs.  
 A detailed analysis of the topological phase and MZMs taking into account the lattice mismatch betwen ZGNR and Ising superconductor, as well as finite size effects with various ZGNRs are presented in the Supplementary Notes 6 and 7 \cite{SM}.

\section{Discussion} 
Remarkable experimental progress has been made in the fabrication of high-quality devices to engineer one-dimensional topological superconductors with MZMs~\cite{Mourik2012,Das2012,Deng2016,Albrecht2016,Fornieri2019}. 
However, many questions still remain about whether the experimental signatures often associated with MZMs are truly unique to these phases and if the experiments have unambiguously detected these modes~\cite{Frolov2020,Yu2021}.  
In general, the difficulties in establishing the presence of MZMs in these devices stem from two main challenges. 
First, the disorder, which is inevitable in even the best quality devices, can obscure the  MZMs. 
This leads to a much smaller region of the phase diagram that is truly topological as compared to simplified models. 
Moreover, the disorder localized trivial subgap states can often give very similar experimental signatures that are considered the hallmark of MZMs~\cite{Yu2021}. This creates an ambiguity in interpreting the experimental data.  
Second, it is often difficult to ensure that the proximity induced superconducting gap is indeed achieved. 
This difficulty is amplified by the fact that, in many systems, a magnetic field is required to reach the topological phase. 
Hence, one needs a delicate balance of the magnetic field, such that it is large enough to achieve the topological phase but small enough to preserve the superconductivity. 
Indeed, this is the case for the semiconducting quantum wires, which are considered among most promising platforms to detect MZMs due to the superior fabrication quality available in semiconductor devices compared to the topological insulators. 

Given that our proposal is based on quasi one-dimensional graphene nanoribbons, we expect that high-quality devices comparable to semiconductor quantum wires are achievable, especially given the typically low concentration of defects that form in atomically precise ZGNRs \cite{Ruffieux2016, Pizzochero2021, Pizzochero2022}. Furthermore, our proposal has the added strength that it is based on an all-electrical setup, requiring only gate control and electric fields to achieve topological phase. 
This can prove to be an important advantage over conventional semiconductor quantum wires and may lie at the sweet spot of low disorder and easy tunability. Moreover, the basic components, ZGNRs and underlying Ising superconductor are both low-dimensional materials, which may facilitate the realization of complex device geometries at the ultimate limit of atomic thinness.

It is important to highlight the assumptions underlying our analysis, which will motivate future studies. First, we have assumed that the $\alpha$-spins of the ZGNR are polarized in the plane of the heterostructure. It is crucial that the half-metal spins have a component orthogonal to the Ising spins. 
In the absence of SO coupling in graphene, the dipolar interactions are the most dominant contributor to the magnetic anisotropy. In a ribbon geometry, the anisotropic part of the dipolar interaction tends to align magnetic dipoles along the ribbon length, thus favoring the in-plane polarization of the magnetic state of the ZGNR. 
Since the most Ising superconductors have out-of-plane spin anisotropy, this satisfies our requirement.
Alternatively, a small magnetic field can be used to ensure in-plane alignment of $\alpha$ spins.
The very high in-plane $H_{c2}$ and strong out-of-plane anisotropy of the Ising superconductor will ensure that superconductivity is unaffected by this magnetic field. 
Recently, an in-plane Ising superconductivity, where the SO coupling locks the spin quantization of the superconductor in the two-dimensional plane was identified in the $\text{EuO}_x$/$\text{KTaO}_3$ (110) interfaces~\cite{Yang2025}, which may constitute a viable system to remedy the possible out-of-plane spin polarization of ZGNRs. Second, we have assumed a perfect matching of the the hexagonal lattices of the Ising superconductor and ZGNR.  
This approximation is introduced for simplicity and is not expected to play an important role in the resulting physics.
Third, we have neglected any interfacial effects. In real systems, the interface further break inversion symmetry, which can induce Rashba SO coupling and allow spin-flip hoppings between the superconductor and GNR. 
We expect this effect to add to the induced superconductivity and increase the odd parity gap~\cite{Chung2011}. 
Finally, we have not considered the effect of disorder.

\section{Conclusion}
To conclude, we have shown that spin-singlet $s$-wave Ising superconductor can induce odd-parity triplet pairing in half metals, as long as the half-metal and Ising spins have different spin-quantization axis. We have identified half-metallic zigzag graphene nanoribbons as a promising platform for realizing one-dimensional topological superconductors hosting Majorana zero modes at their ends. 
Compared to existing platforms -- such as semiconductor nanowires, topological insulators, and ferromagnetic atom chains -- half-metallic nanoribbons offer significant advantage, including fully electrical operation without an external magnetic field, low defect densities, and a two-dimensional nature, making them a promising platform for realizing complex quantum gate architectures.

\vspace{\baselineskip}

%
%

\ack{G.C. acknowledges support from Engineering and Physical Sciences Research Council (EPSRC) grant EP/V062654/1 for the work done in Cambridge.}




\bibliography{GNR_Majorana_bibliography}

\end{document}


\title{Supplemental Material for ``Zigzag Graphene Nanoribbons as a Majorana Platform"}
\author{Ruize Ma}
\affiliation{Department of Physics, ETH Z\"urich, Zurich 8093, Switzerland}

\author{Michele Pizzochero}
\email{mp2834@bath.ac.uk}
\affiliation{Department of Physics, University of Bath, Bath BA2 7AY, United Kingdom}
\affiliation{School of Engineering and Applied Sciences, Harvard University, Cambridge, MA 02138, United States}

\author{Gaurav Chaudhary}
\email{gchaudhary0806@gmail.com}
\affiliation{TCM Group, Cavendish Laboratory, University of Cambridge, 
Cambridge CB3 0US, United Kingdom}
\affiliation{School of Physics, University of Bristol, Tyndall Avenue, Bristol BS8 1TL, United Kingdom}

\maketitle




\appendix
\onecolumngrid
\tableofcontents
\renewcommand\theequation{S\arabic{equation}}
\renewcommand\thefigure{S\arabic{figure}}
\setcounter{equation}{0}
\setcounter{figure}{0}



\newpage
\subsection{Supplementary Note 1: Perturbation theory derivation of induced pairing}
Starting from Eqs. 1-2 of the main text, we wish to eliminate the coupling term Eq.2 (c) and obtain the effective Hamiltonian for the half metal. Below, we accomplish this using Schrieffer-Wolff transformation:
\begin{align}\label{Eq:Supp_SW1}
    \tilde{H} = \text{e}^{S} H \text{e}^{-S} .
\end{align}
By treating weak tunnel coupling as a perturbative term, we can cast the transformed Hamiltonian as
\begin{align}\label{Eq:Supp_SW2}
    \tilde{H} = H_{HM} + H_{SC} + \frac{1}{2}[S, H_T] + O(H^3_T).
\end{align}
Here, we have chosen the transformation $S$, such that
\begin{align}\label{Eq:Supp_SW_constraint}
    H_T + [S, H_{HM} + H_{SC}] = 0,
\end{align}
which is always possible. The third term on the R.H.S of Eq.~\ref{Eq:Supp_SW2} is the second-order perturbation correction that we wish to calculate. 
Consider the transformation $S$ to take the form
\begin{align}\label{Eq:Supp_transform}
    S = \sum_{k,q} \sum_{\sigma} \alpha_{\sigma, s}(k,q) \hat{\chi}_{k,s} \hat{\gamma}_{q,\sigma} +  \beta_{\sigma,s}(k,q) \hat{\chi}^{\dagger}_{k,s} \hat{\gamma}_{q,\sigma} - \text{h.c.}
\end{align}
Notice we have explicitly constrained $S$ to be anti-Hermitian as required for the transformation $\exp(S)$ to be unitary. 
The second-order perturbation correction can be calculated after some cumbersome but straightforward manipulation of the commutators
\begin{align}\label{Eq:Supp_SW_decouple}
    & [S,\, H_T] =  \sum_{k, k' } \sum_{\sigma, \sigma'} \alpha_{\sigma,s}(k,k') [ T^{\ast}_{\sigma,s}(k') \hat{\chi}_{k,s} \hat{\chi}_{k',s} - T_{\sigma',s}(k') \hat{\gamma}_{k,\sigma'} \hat{\gamma}_{k',\sigma}  ] + 
    \beta_{\sigma,s}(k,k') [ T^{\ast}_{\sigma,s}(k') \hat{\chi}^{\dagger}_{k,s} \hat{\chi}_{k',s} - T^{\ast}_{\sigma',s}(k) \hat{\gamma}_{k,\sigma'} \hat{\gamma}_{k',\sigma}  ] \notag\\ 
    & \hspace{3cm}+ \text{h.c.}
\end{align}
From the R.H.S above, it is clear that up to the second order in perturbation, our choice of $S$ eliminates the coupling term between the half-metal and the superconductor, while introducing pairing terms in the half-metal and renormalizing the dispersions of half-metal and superconductors. Our goal is to find the coefficients $\alpha$ and $\beta$ using the constraints imposed by Eq.~\ref{Eq:Supp_SW_constraint}. For this part, after some algebra, we obtain  
\begin{subequations}\label{Eq:Supp_SHCommut}
\begin{align}
    & [S, \, H_{HM}] = \sum_{k,k'}\sum_{\sigma} \epsilon(k) [\alpha_{\sigma,s} (k,k') \hat{\chi}_{k,s} \hat{\gamma}_{k',\sigma} - \beta_{\sigma, s}(k,k') \hat{\chi}^{\dagger}_{k,s} \hat{\gamma}_{k',\sigma} ] + \text{h.c.} \\ 
    & [S, \, H_{SC}] = \sum_{k,k'}\sum_{\sigma} \xi(k) [\alpha_{\sigma,s} (k,k') \hat{\gamma}_{k',\sigma} \hat{\chi}_{k,s} + \beta_{\sigma, s}(k,k') \hat{\chi}^{\dagger}_{k,s} \hat{\gamma}_{k',\sigma} ] + 2 \sigma \Delta [\alpha_{\sigma,s}(k,k') \hat{\gamma}^{\dagger}_{-k',-\sigma} \hat{\chi}_{k,s} + \beta_{\sigma,s}(k,k') \hat{\chi}^{\dagger}_{k,s}\hat{\gamma}^{\dagger}_{-k',-\sigma} ]\notag\\
    &\hspace{3cm} + \text{h.c.} 
\end{align}
\end{subequations}
Here the spin $\sigma = \uparrow (\downarrow)$, when appears as a coefficient in the equation, is interpreted as $1 (-1)$. 
Plugging Eq.~\ref{Eq:Supp_SHCommut} into Eq.~\ref{Eq:Supp_SW_constraint}, we obtain
\begin{align}\label{Eq:Supp_plug_back}
    & \sum_{k,k'}\sum_{\sigma} [T_{\sigma,s}(k) \delta_{k,k'} + \beta_{\sigma,s}(k,k')\{ \xi(k) - \epsilon(k)\} - 2\sigma\Delta \alpha^{\ast}_{-\sigma,s}(k, -k')] \hat{\chi}^{\dagger}_{k,s} \hat{\gamma}_{k',\sigma} \notag\\
    & \hspace{1cm} + [ \alpha_{\sigma,s}(k,k')\{ \epsilon(k) - \xi(k)\} - 2\sigma\Delta \beta^{\ast}_{-\sigma,s}(k,-k')  ] \hat{\chi}_{k,s} \hat{\gamma}_{k',\sigma} + \text{h.c.} = 0
\end{align}
Setting coefficient of each independent set of Fermion bilinear to zero
\begin{subequations}\label{Eq:Supp_alphabeta}
\begin{align}
    & \alpha_{\sigma,s}(k,k') = -\delta_{k,-k'} \frac{2\sigma \Delta T^{\ast}_{-\sigma,s}(k)}{4\Delta^2 - [ \epsilon(k) - \xi(k)]^2} ,\label{Eq:Supp_alpha} \\
    & \beta_{\sigma,s}(k,k') = -\delta_{k,k'} \frac{T_{\sigma,s}(k) [\epsilon(k) - \xi(k)]}{4\Delta^2 - [ \epsilon(k) - \xi(k)]^2} . \label{Eq:Supp_beta}
\end{align}
\end{subequations}
As expected from our initial assumption of momentum conserving process, the pairing encoded in $\alpha$ coefficients is only non-zero for $k=-k'$ and band renormalization encoded in $\beta$ coefficients is only non-zero for $k=k'$. 
We obtain the induced pairing
\begin{align}\label{Eq:Supp_induced_gap}
    \tilde{\Delta}(k) = 2\Delta\frac{T_{\uparrow,s}(k) T_{\downarrow,s}(-k) - T_{\downarrow,s}(k) T_{\uparrow,s}(-k)}{4\Delta^2 - [\epsilon(k) - \xi(k)]^2} .
\end{align}
Since the above expression has divergence, it can incorrectly indicate that very large induced pairings are possible. However, notice that as the induced pairing term reaches divergence, other second order perturbation corrections, such as correction to the band dispersion also diverge at the same rate. Therefore, corrections only up to second order are no longer valid in this limit, and one needs to consider higher order terms. 

\newpage
\subsection{Supplementary Note 2: Finite induced gap from Ising spin-orbit coupling}
To analytically show that induced gap is finite when superconductor has Ising spin-orbit coupling, we consider the two-orbital model for the normal state of the superconductor presented in Eq. 6 of the main text . The numerator of Eq.~\ref{Eq:Supp_induced_gap} contains the crucial combination of the band projected tunnelings that need to be finite, which can be written as
\begin{align}\label{Eq:Supp_Proj_terms}
    T_{\uparrow,\rightarrow}(k) T_{\uparrow,\rightarrow}(-k) - T_{\downarrow,\rightarrow}(k) T_{\uparrow,\rightarrow}(-k) = - 
    \frac{1}{2} \sum_{\sigma} \sigma \biggl ( \sum_{a,b} U^{i,a}(k) t_{b} V^{\sigma\ast}_{j,b}(k) \biggr ) \biggl ( \sum_{a,b} U_{i,a}(-k) t_{b} V^{-\sigma\ast}_{j,b}(-k) 
    \biggr )
\end{align}
Here $U$ and $V$ are the unitary transformations that diagonalize the half-metal and the normal state of the Ising superconductor from a microscopic tight binding Hamiltonian, such that $i$ and $j$ are respective band labels, $a$ and $b$ are respective orbital labels, and $t_b$ are the tunneling amplitudes obtained by orbital overlap. 
For the two-orbital model in Eq. 6 of the main text, the SO coupling does not mix the spin, in the sense that the Hamiltonian is block diagonal in the spin and band eigenstates can be labeled by their spins. Indeed, the two opposite spin  blocks are symmetric under $H^{\ast}_{I,\sigma}(-k) = H_{I,-\sigma}(k)$. Nevertheless, the SO coupling breaks $SU(2)$ symmetry. The inversion symmetry is broken in the sense that $E_{\sigma}(k) \neq E_{\sigma} (-k)$. Inversion symmetry in band dispersion is restored as $E_{\sigma}(k) = E_{-\sigma} (-k)$. This is exactly manifested as the spin-valley locking in Ising systems when the full two-dimensional Hamiltonian is considered. 

Each of the spin block of the Ising metal can be diagonalized under the transformation
\begin{align}
    V_{\sigma}(k) = \cos\frac{\theta_{\sigma}(k)}{2} \tau_3 & + \sigma \sin\frac{\theta_{\sigma}(k)}{2} \tau_2 .
\end{align}
Here $\theta_{\sigma}$ is the azimuthal angle after the standard mapping of $2\times 2$ Hermitian matrices to the Bloch sphere. 
The above parameterization satisfies the following conditions due to the underlying crystalline symmetries
\begin{subequations}
\begin{align}
    & \cos\frac{\theta_{\sigma}(k)}{2} = \cos\frac{\theta_{-\sigma}(-k)}{2},\\
    & \sin\frac{\theta_{\sigma}(k)}{2} = \sin\frac{\theta_{-\sigma}(-k)}{2}.  
\end{align}
\end{subequations}
Crucially, because
\begin{align}
    \sin\frac{\theta_{\sigma}(k)}{2} \neq \sin\frac{\theta_{\sigma}(-k)}{2}, \quad 
    \cos\frac{\theta_{\sigma}(k)}{2} \neq \cos\frac{\theta_{\sigma}(-k)}{2},
\end{align}
as long as the SO term $\lambda$ is non-zero, in the evaluated expression for Eq.~\ref{Eq:Supp_Proj_terms}, we get non-zero contributions that go as
\begin{align}\label{Eq:S_sin}
    \sin\frac{\theta_{\sigma}(k)}{2} - \sin\frac{\theta_{\sigma}(-k)}{2}
\end{align}
and 
\begin{align}\label{Eq:S_cos}
    \cos\frac{\theta_{\sigma}(k)}{2} - \cos\frac{\theta_{\sigma}(-k)}{2},
\end{align}
which ensure a finite proximity gap.

Here we also clarify that the pairing used in the main text tight binding model is a spin-singlet and even parity. While the pairing gap in the main text is written in a tight-binding basis, since the conventional BCS superconductivity is a Fermi surface instability the true pairing symmetry of the gap should be analyzed by considering the gap symmetry in the band basis. Projection of the pairing term in Eq.~6 of the main text from the tight-binding to the band basis can be written as follows:
\begin{align}
    &\sum_{\tau,\sigma} \Delta \sigma \hat{f}^{\dagger}_{\tau\sigma}(k) \hat{f}^{\dagger}_{\tau,-\sigma}(k) = 
     \Delta  \biggl (\cos \frac{\theta_{\uparrow}(k)}{2} \cos \frac{\theta_{\downarrow}(k)}{2} + \sin \frac{\theta_{\uparrow}(k)}{2} \sin \frac{\theta_{\downarrow}(k)}{2} \notag\\ 
    &\hspace{4cm}+i \cos \frac{\theta_{\uparrow}(k)}{2} \sin \frac{\theta_{\downarrow}(k)}{2} 
    -i \sin \frac{\theta_{\uparrow}(k)}{2} \cos \frac{\theta_{\downarrow}(k)}{2} \biggr ) \sum_{\sigma} \sigma\gamma^{\dagger}_{\sigma}(k) \gamma^{\dagger}_{-\sigma}(k)
\end{align}
Using the relation of Eqs.~\ref{Eq:S_sin} and \ref{Eq:S_cos} above, we can conclude that RHS is even in $k$ and its spin structure is purely singlet.

\newpage
\subsection{Supplementary Note 3: Tight binding results for GNR electronic structure}

\begin{figure}[ht]
    \centering
    \includegraphics[width=0.92\linewidth]{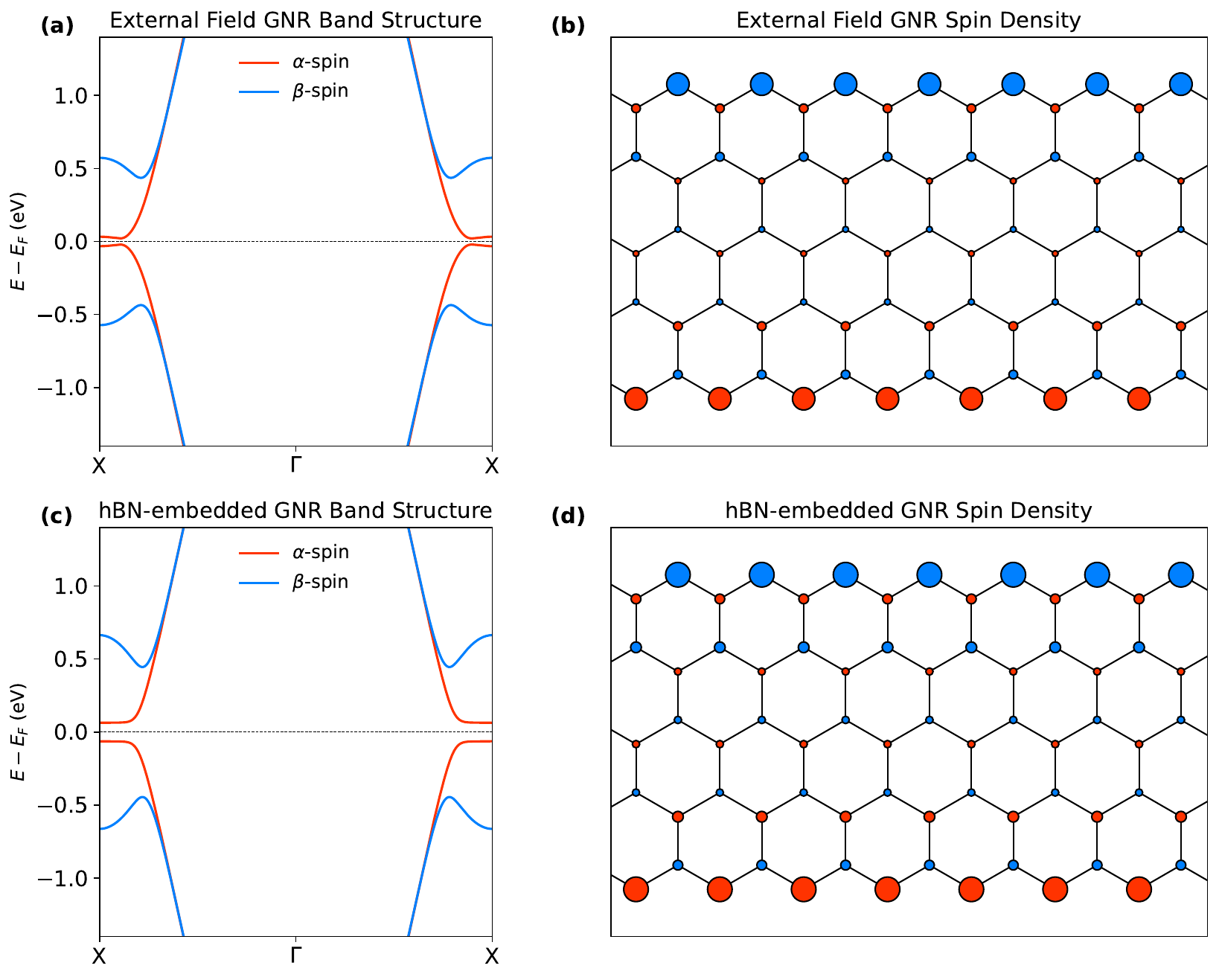}
    \caption{Equivalence of field-induced and embedding hBN-induced ZGNR magnetic edge states. (a) The electronic band structure of ZGNR in transverse in-plane electric field $F=0.07\ \text{V/nm}$. (b) The corresponding spin density plot ($n_\alpha - n_\beta$), where the color indicates the majority spin at each site, and the size indicates the magnitude of the spin density difference. (c, d) The band structure and spin density plot of ZGNR embedded in hBN, which is represented by onsite potentials $\epsilon_1 = -\epsilon_2 = 0.3\ \text{eV}$ at the two edges. The two cases clearly show same magnetic edge states and will thus exhibit identical topological properties when proximitized with an Ising superconductor.}
    \label{fig:sm_density}
\end{figure}

Here we present more results on the electronic structure and spin densities of the ZGNRs to illustrate the edge magnetism and half metallic features. 
In Fig.~\ref{fig:sm_density}, we present the tight-binding band structure and real-space spin density plots for both the ZGNR coupled with an external in-plane electric field and ZGNR embedded within the hBN. 
In both cases, the results look remarkably similar.
On the tight-binding level, the hBN can be represented by onsite potentials $\epsilon_1 = -\epsilon_2$ localized at the opposite edges of the GNR respectively. This model is shown to reproduce the DFT results\cite{Tepliakov2023}. Comparing the two methods for obtaining the magnetic edge states, we see that the resulting states are very similar. If we couple both of them to Ising superconductor proximately, the resulting topological phase transitions would be expected to be very similar to Fig.~2 in the main text.

\newpage
\subsection{Supplementary Note 4: A simple model}
To better understand the results of the main text, we also develop a simple model that describes the minimal low-energy electronic features of antiferromagnetic and half-metallic  ZGNRs. We consider  two coupled spinful diatomic chains
\begin{align}\label{Eq:DiChain}
    & H_{0}(k) = -\mu + M s_3 \eta_3  + \frac{\gamma}{2} \eta_3 + \frac{\lambda}{2} (\eta_3 + \tau_3) +  w  \eta_1 + (t+t'\cos k)\tau_1 + t'\sin k \tau_2 ,
\end{align}
where $s, \eta,$ and $\tau$ Pauli matrices act in spin, edge, and sublattice spaces,  respectively. By envisaging the two chains as the two edges of the ZGNR, $t,\, t'$ are the nearest neighbor hoppings within each zigzag edge, $w$ is the inter-edge tunneling, $M$ is the on-site magnetic moment that flip the sign at opposite edges, in line with the edge antiferromagnetic state, $\gamma$ represents a transverse in-plane electric field that breaks left and right edge degeneracy, and $\lambda$ represents an on-site sublattice potential inherited from  embedding in hexagonal boron nitride.
With the freedom to tune the individual parameters in the model, we obtain a low-energy band structure of the ZGNR that is in a good agreement with the microscopic tight-binding and first-principles calculations. To match the low energy sector of this model with the microscopic calculations, however, the actual values of these parameters are not the same as the analogous terms in the microscopic model. In what follows, we restrict our investigation to this analytical model, which can be understood in terms of a mapping to the Kitaev chain model~\cite{Kitaev2001}. 

\begin{figure}[ht]
    \centering
    \includegraphics[width=0.92\linewidth]{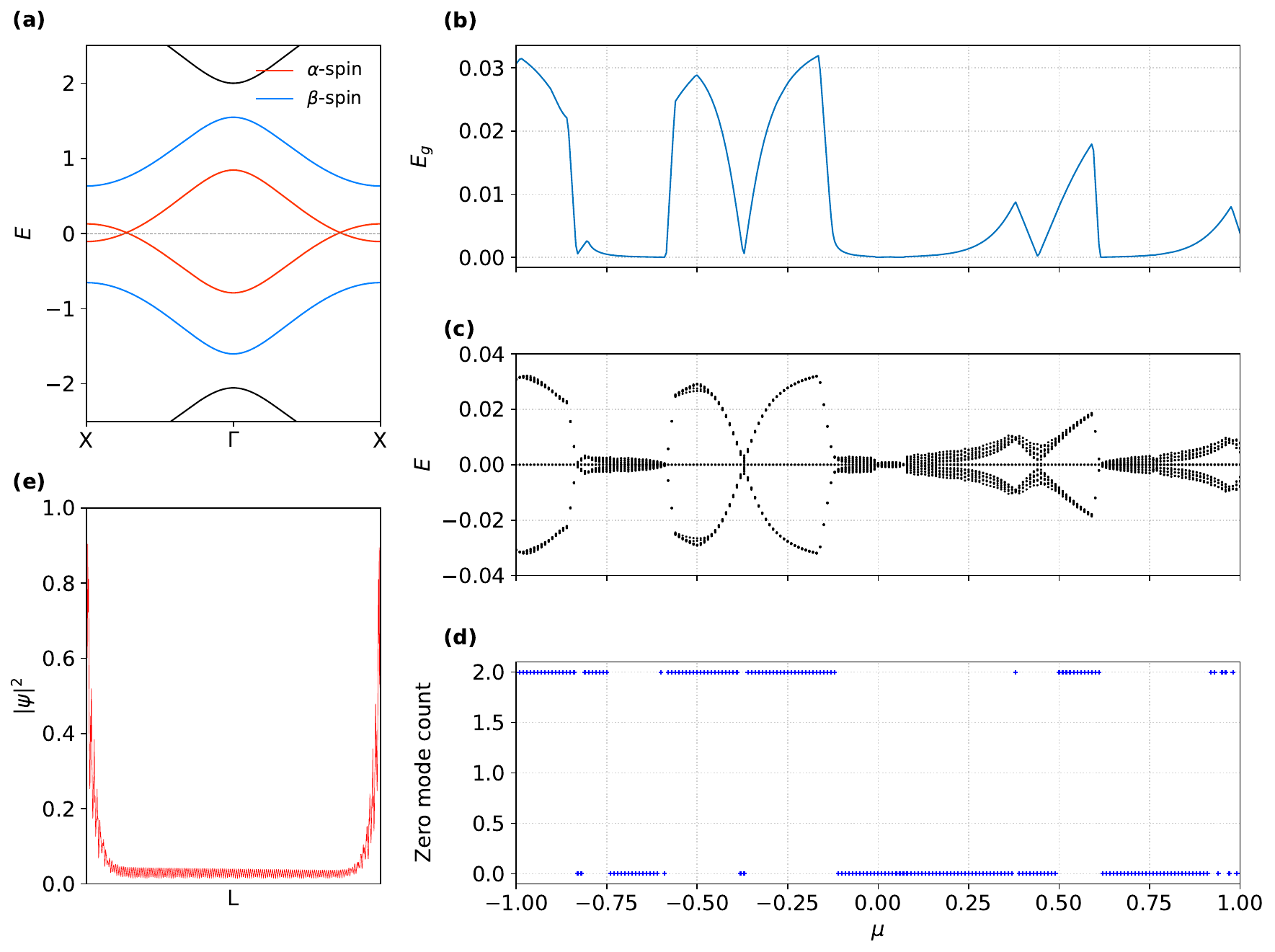}
    \caption{\label{fig:Diatomic}
  Calculations for the simplified diatomic chain model. We have taken model parameters as $t = 1,\, t' = 0.5,\, w = 0.1,\, \gamma = 0,\, M = 0.82, \, \lambda = 0.26$ in the $H_0$, and $t_0 = 0.67,\, t_2 = 1.1,\, t_3 = 0.47,\, \lambda' = 0.2$ in $H_I$, pairing gap $\Delta = 0.04$, and the tunnel coupling between half-metal and Ising SC is 0.2.
  (a) The bandstructure of the diatomic chain with half (semi)metallic features, (b) Bogoliubov excitation gap for the coupled system with periodic boundary condition shows clear gap in the spectrum, (c) Bogoliubov excitation spectrum with open boundary condition shows zero energy modes separated from the rest of the spectrum by the bulk gap, (d) Number of zero energy mode in the finite size system, and (e) probability distribution of zero energy modes over the system at $\mu = -0.25$.}
\end{figure}

In Fig.~\ref{fig:Diatomic}(a), we show the band structure of the diatomic chain in the absence of the coupling to the superconductor. At charge neutrality, it consists of a pair of $\alpha$-spin bands crossing in the vicinity the $X$-point along with gapped $\beta$-spin bands. The model parameters are chosen to obtain an agreement with Fig.~\ref{fig:Diatomic}(c).
In Fig.~\ref{fig:Diatomic} (b), we show the Bogoliubov excitation gap of the proximity coupled system as a function of the Fermi level in the diatomic chain, as varied within the band width of the $\alpha$-spin bands.  
As expected from our main result, we observe that the system is usually gapped, showing that the spin-singlet Ising superconductor is indeed effective to induce a proximity pairing gap in the half-metal even when the two systems are coupled by a simple elastic tunneling process, without invoking any spin-flip or interfacial Rashba effects.
The spectrum of $16$ Bogoliubov excitations closest to zero energy in a representative finite system of $250$ unit cells is given in Fig.~\ref{fig:Diatomic}(c), For the most part when $\mu$ is within the $\alpha$ spin band, the system exhibits zero-energy modes that are separated from the rest of the spectrum by the bulk gap, as expected for this one-dimensional topological phase. 
In Fig.~\ref{fig:Diatomic}(d), we show the count of zero-energy modes (within $10^{-8}$ numerical accuracy). In the whole region under consideration, there are either no zero modes, corresponding to the trivial phase, or exactly two zero models (i.e., one mode at each end of the ZGNR), thus ensuring that the zero modes are well isolated. This is further supported from the visualization of the wavefunction along the system in In Fig.~\ref{fig:Diatomic}(e), which localizes at the opposite ends of the ZGNR and rapidly decays into the bulk.

The findings overviewed in Fig.~\ref{fig:Diatomic} can be understood analytically by restricting the investigation to the two $\alpha$-spin bands,
\begin{align}\label{Eq:LowBand}
&\epsilon_{\alpha\pm}(k) = \pm \sqrt{\biggl (M+\frac{\lambda+\gamma}{2} \biggr)^2 + w^2} \mp \sqrt{\frac{\lambda^2}{4} + t^2 + t'^2+2tt'\cos k},    
\end{align}
which is an even function of $k$. 
When $\mu$ intersects only the lower band $\epsilon_{\alpha,-}$, the system is in the topologically non-trivial phase, while any pairing in $\epsilon_{\alpha,+}$ band is in the trivial limit.  As a result, isolated MZMs form at the two ends of a finite system.  As $\mu$ approaches charge neutrality, it intersects both $\epsilon_{\alpha,\pm}$ bands. Hence, both of them are in the topologically non-trivial phase. 
In this case, since each end hosts two MZMs, the modes hybridize to give a complex fermion and the system is overall trivial with no MZMs. 
As $\mu$ is further increased to only reside in $\epsilon_{\alpha,+}$ band, the topologically non-trivial phase is restored.
In this region, the induced gap is smaller and the MZMs at opposite ends have a small hybridization gap due to the finite-size effect. 
Therefore, in Fig.~\ref{fig:Diatomic}(d), modes with numerical values below the threshold of $10^{-8}$ are not included, resulting in a zero count within this window.

\newpage
\subsection{Supplementary Note 5: Scalling with Pairing and Interlayer tunneling}
In this note, we study the proximity induced gap in the ZGNRs and its scaling with the gap $\Delta$ of the parent Ising superconductor and interlayer tunneling.
We study the main-text width-three zigzag GNR coupled to a three-row, two-orbital Ising superconducting strip. The parameters are $t=U=2.88$ eV, $F=0.12\,\mathrm{V}/\text{\AA}$ and $|w_{\tau}|=0.10$ eV. Chemical potentials use the plotted convention $\mu_b=\mu+1.44$ eV. Remaining Hamiltonian parameters and prefactors follow the main-text model. The GNR Hartree background is fixed for the superconducting spectra.

\begin{figure}[ht]
    \centering
    \includegraphics[width=0.94\linewidth]{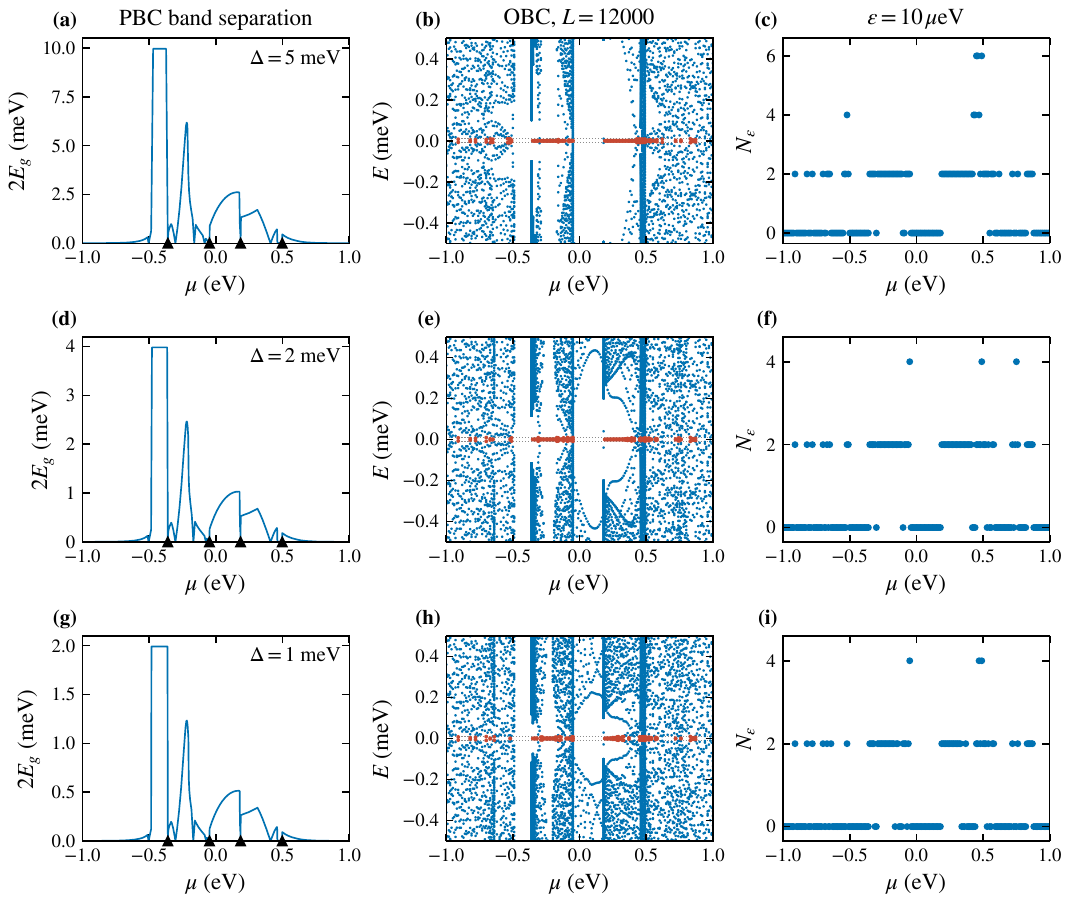}
    \caption{\label{fig:VaryDelta} Dependence on $\Delta=5,2,1$ meV. Left: BdG quasiparticle gap $2 E_g=2\min_{k,n}|E_n(k)|$ under PBC. Triangles mark directly evaluated bulk gap closings. Centre: all OBC eigenvalues in the displayed interval $|E|<0.5$ meV at $L=12000$. Right: Count of approximately zero energy states (i.e. states in OBC that lie in the energy window $\leq 10 \mu\text{eV}$ 
    In the central column, the red spectral points lie in this window; grey dotted lines mark its boundaries.}
\end{figure}
 OBC results use 201 chemical potentials with spacing $0.01$ eV. The PBC curves use adaptive meshes of 663, 550 and 473 points, respectively, including the directly evaluated closings; the maximum independently checked midpoint interpolation error is $1\,\mu$eV. Smaller parent pairing preserves intervals with a low-energy pair while narrowing the excitation gap. The count includes both particle-hole partners and can also include additional small-gap excitations.

\newpage
Now we analyze the scaling of the induced gap at a fixed $\mu = -0.25 eV$, which is near the peak of the topological gap. We find that the induced bulk gap is approximately proportional to the parent pairing when the parent gap $\Delta$ is varied in the range $0.1$--$5$ meV. Its weak-tunneling dependence is close to quadratic. 
These scaling dependences are also consistent with our perturbation theory expression in Eq. (4) of the main text.
A smaller gap increases the length needed to separate the end states in energy and space.
\begin{figure}[ht]
    \centering
    \includegraphics[width=0.94\linewidth]{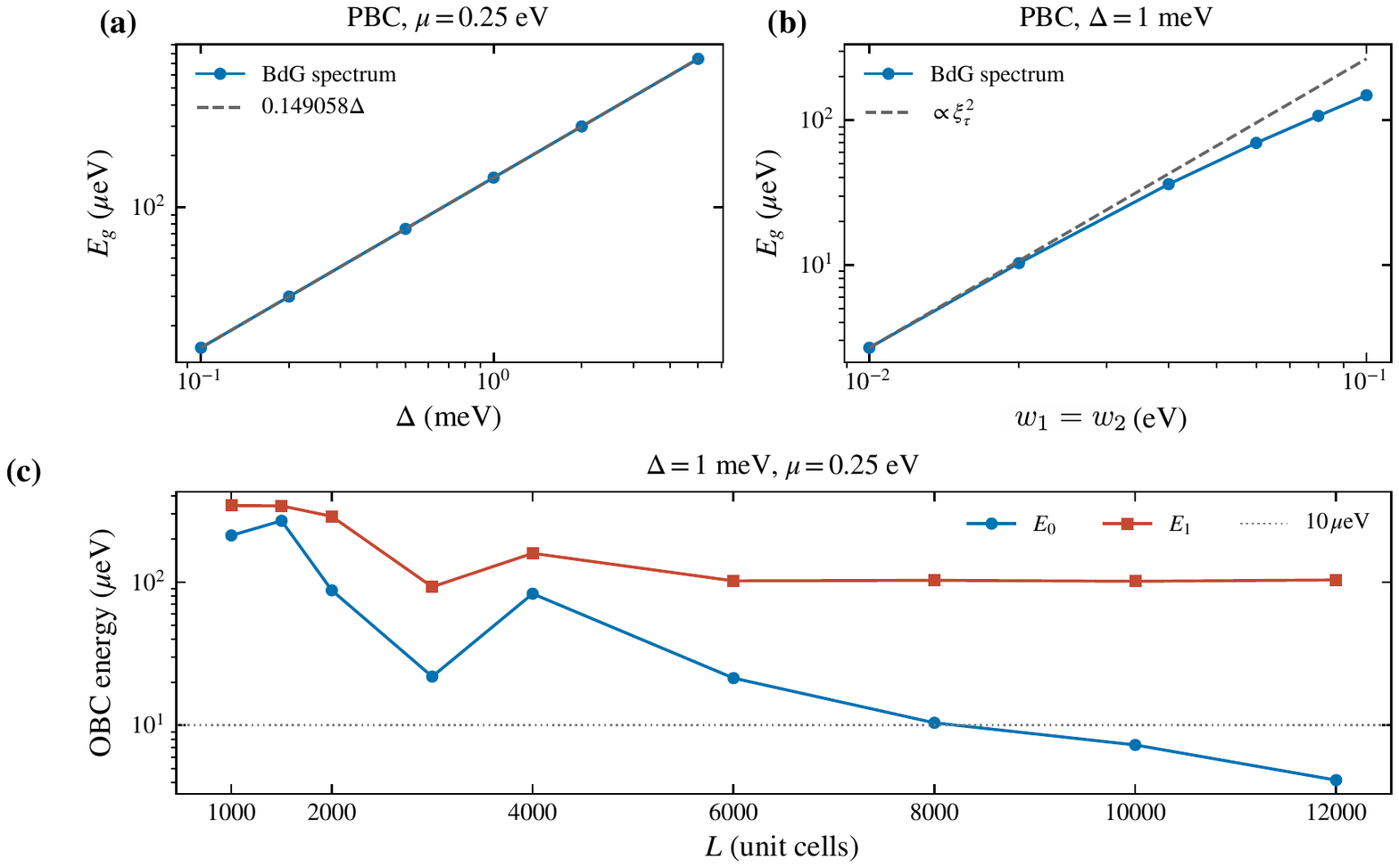}
    \caption{\label{fig:scalling} (a) Bulk $E_g$ versus $\Delta$ at $w_\tau=0.10$ eV; $E_g/\Delta$ varies from $0.149058$ to $0.148805$. (b) Tunnelling dependence at $\Delta=1$ meV. The dashed quadratic reference describes the weak-coupling trend; the initial doubling gives a logarithmic slope of $1.945$. (c) Lowest and next positive OBC energies, $E_0$ and $E_1$, versus length at $\Delta=1$ meV and $\mu=-0.25$ eV. At $L=12000$, $E_0=4.144\,\mu$eV, $E_1=103.58\,\mu$eV and $E_g=149.05\,\mu$eV. The counting window lies between the split pair and the next excitation. Finite-length overlap can produce oscillatory splitting as length changes~\cite{DasSarma2012}. The appropriate convergence length therefore depends on the gap and chemical potential, rather than on the parent pairing alone.}
\end{figure}

\newpage
\subsection{Supplementary Note 6: Finite size effects and delocalisation of end states}

\begin{figure}[ht]
    \centering
    \includegraphics[width=0.97\linewidth]{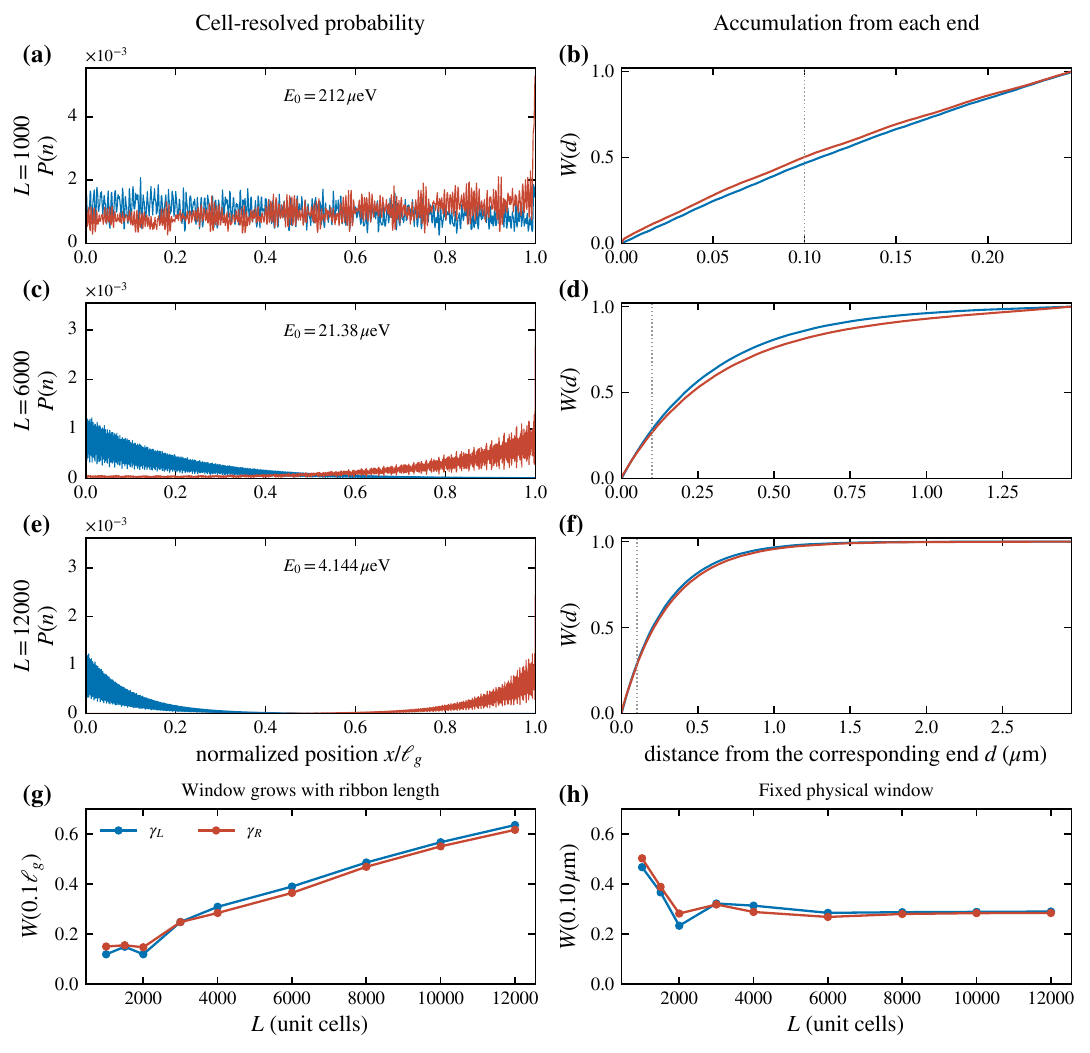}
    \caption{\label{fig:Endstates} Spatial evolution at $\Delta=1$ meV and $\mu=-0.25$ eV. (a,c,e) Cell-resolved probabilities of $\gamma_L$ (blue) and $\gamma_R$ (red) at $L=1000,6000,12000$. Position is normalized by $\ell_g=(L-1)a_g$, with $a_g=\sqrt3(1.42\,\text{\AA})$. (b,d,f) Probability accumulated from the corresponding end within physical distance $d$; the dotted line marks $0.10\,\mu$m. (g) Weight in the outermost 10\% of cells. (h) Weight within a fixed $0.10\,\mu$m of each end.}
\end{figure}

For the lowest particle-hole pair,
\[
\gamma_L=\frac{e^{i\varphi}\psi_++e^{-i\varphi}\mathcal C\psi_+}{\sqrt2},\qquad
\gamma_R=\frac{-i(e^{i\varphi}\psi_+-e^{-i\varphi}\mathcal C\psi_+)}{\sqrt2},\qquad \mathcal C=\tau_xK,
\]
with the phase maximizing the left-end weight. These normalized self-conjugate components describe the lowest two-state subspace; they remain coupled at finite length. The overlapping short-chain profiles develop distinct peaks at opposite ends. The outermost-10\% window grows with length, whereas panel (h) uses a fixed physical distance.

\newpage
\subsection{Supplementary Note 7: Lattice mismatch and geometry dependent interlayer tunneling}
In this note we study the dependence on the lattice mismatch between the ZGNR and Ising superconductor by relaxing the constraint used in the main text, which put each triangular lattice site of the metal right under a carbon of the ZGNR. 
Now, the SC still retains the two metal-centred effective orbital channels of the main-text model. Both carbon sublattices are eligible to tunnel; actual contacts are selected by carbon--metal separation. Orbital selection in the substrate is distinct from the lateral registration of carbon atoms~\cite{Koshino2015, Gani2019}.
\begin{figure}[ht]
    \centering
    \includegraphics[width=0.88\linewidth]{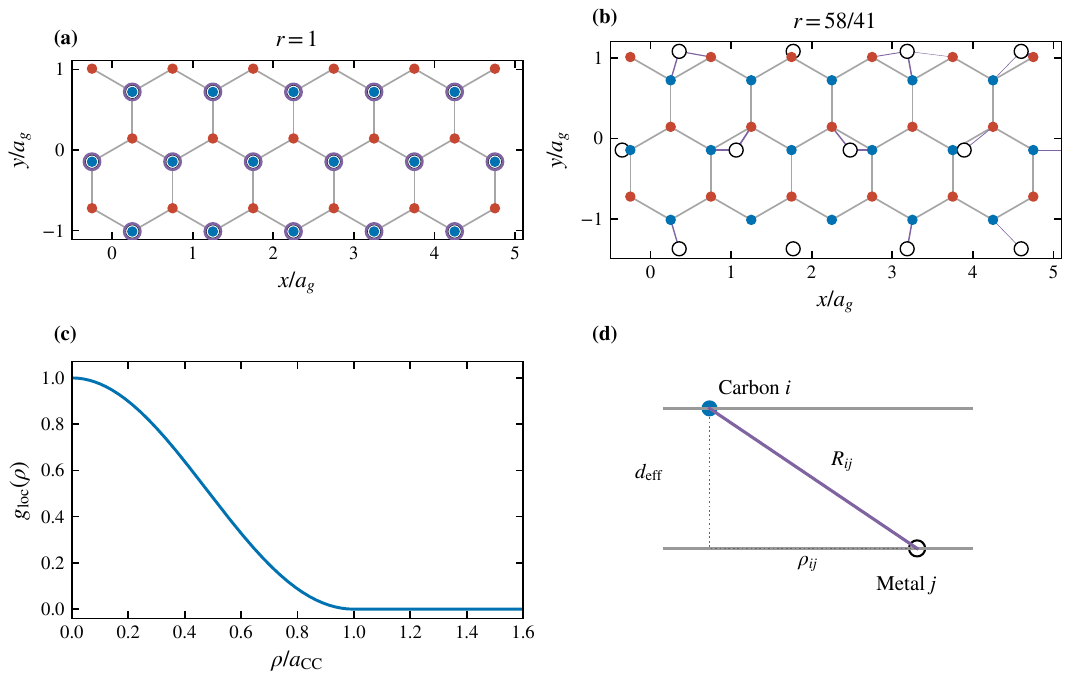}
    \caption{\label{fig:geom} (a,b) Coordinate-generated top views at matched and isotropically mismatched registration. Red and blue circles denote carbon A and B sites; grey bonds connect carbon neighbours. Hollow circles mark SC metal centres. Purple rings in (a) indicate coincident B--metal contacts, while purple links in (b) show the nonzero distance-selected contacts, with thickness set by their coupling. (c) Compact contact envelope. (d) Effective distance construction, $R_{ij}=\sqrt{d_{\rm eff}^2+\rho_{ij}^2}$}
\end{figure}

Keeping the GNR fixed, the SC coordinates are
\[
\mathbf R_j^s=\operatorname{diag}(r_x,r_y)
 (\widetilde{\mathbf R}_j^s-\mathbf R_0^s)+\mathbf R_0^g+\mathbf u,
\qquad \rho_{ij}=|\mathbf R_i^g-\mathbf R_j^s|.
\]
The centres refer to the GNR B sites and the reference three-row SC core. We use $r_x=r_y=r$, $\mathbf u=0$, and
\[
T_{ij}^{(\tau)}=\xi_\tau g_{\rm loc}(\rho_{ij})
\begin{pmatrix}1&1\\1&-1\end{pmatrix},\qquad
g_{\rm loc}(\rho)=
\begin{cases}
\displaystyle e^{-(\sqrt{d_{\rm eff}^2+\rho^2}-d_{\rm eff})/\ell}
 \left(1-\frac{\rho^2}{r_c^2}\right)^2,&0\le\rho<r_c,\\[3pt]
0,&\rho\ge r_c.
\end{cases}
\]
Here $i$ labels carbon, $j$ a metal centre, and $\tau=1,2$ the colocated orbitals. Rows use GNR spins $(\alpha,\beta)$ and columns SC spins $(\uparrow,\downarrow)$. The normalized spin rotation contains $1/\sqrt2$; that normalization is already absorbed in the main-text $\xi_\tau$ entry convention.

The radial exponential follows the distance-decay form used in Slater--Koster interlayer models [2,4]. We use $r_c=a_{\rm CC}=1.42$ \AA, $d_{\rm eff}=3.57$ \AA\ and $\ell=0.184a_g\simeq0.453$ \AA, where $a_g=\sqrt3a_{\rm CC}$. In the graphene parameterization of Ref.~\cite{Moon2013}, the decay length is chosen so that the next-nearest intralayer hopping is $10\%$ of the nearest-neighbour hopping:
\[
e^{-(a_g-a_{\rm CC})/\ell}=0.1
\quad\Rightarrow\quad
\ell=\frac{a_g-a_{\rm CC}}{\ln10}\simeq0.184a_g.
\]
This carbon-based value supplies a phenomenological reference, rather than a fitted carbon--TMD-metal decay length. The compact polynomial is our local-contact assumption: it switches the coupling smoothly to zero at $r_c$, with $g_{\rm loc}(0)=1$ and $g_{\rm loc}(r_c)=g'_{\rm loc}(r_c)=0$. It is separate from the literature exponential, and is not a negligible-tail numerical truncation.

Distances in Fig.\ref{fig:geom} are in \AA\ and $g_{\rm loc}$ is dimensionless. The effective height $d_{\rm eff}$ is a reference decay scale, not a crystallographic carbon--metal height; the interlayer separation reported for graphene/NbSe$_2$ [3] motivates its scale. Throughout, $L$ counts longitudinal GNR cells, with span $\ell_g=(L-1)a_g$. The centred SC strip uses $L_s=1+\operatorname{round}[(L-1)/r]$ cells, approximating the same physical span; $L_s=L$ at match.

At registered match, the SC metal centres coincide with GNR B sites. The nearest A--metal distance is $a_{\rm CC}$, and all noncoincident B--metal distances are at least $a_g$. Hence
\[
g_{\rm loc}(\rho_{ij})=
\begin{cases}1,&\text{coincident B--metal pair},\\0,&\text{every other pair},\end{cases}
\quad\Longrightarrow\quad
H_{\rm loc}^{\rm OBC}(r=1)=H_{\rm main}^{\rm OBC}.
\]
This identity holds for any length and pairing with the same isolated-layer operators and Hartree background. It follows from geometry and the envelope, without a separate matched-case construction. 
\begin{figure}[ht]
    \centering
    \includegraphics[width=1.0\linewidth]{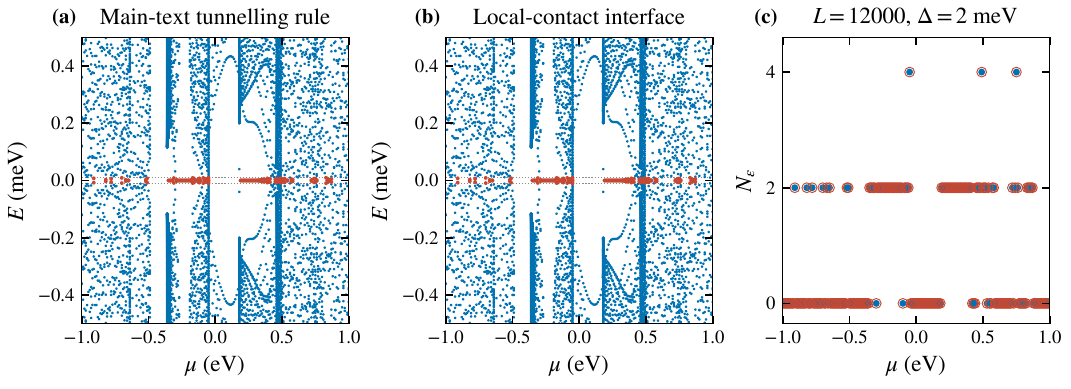}
    \caption{\label{fig:matched} Matched recovery at $L=12000$, $\Delta=2$ meV. (a) Main-text vertical tunnelling rule. (b) Geometrically selected local-contact interface. Both panels show every eigenvalue in $|E|<0.5$ meV at 201 chemical potentials in $[-1,1]$ eV. (c) Corresponding counts for $\varepsilon=10\,\mu$eV: local contact (blue dots) and main-text rule (red open circles). Equality of the complete $576000\times576000$ sparse Hamiltonians is checked coefficient by coefficient under the explicit basis transformation, including the chemical-potential and pairing generators. The resulting operators, spectra and counts coincide throughout the scan.}
\end{figure}

PBC is checked independently against the direct-$k$ main-text Hamiltonian, accounting for the coordinate-derived Bloch gauge and global antiferromagnetic spin orientation. Across 342 control matrices, the maximum matrix difference is $2.2\times10^{-15}$ eV and the maximum complete-spectrum difference is $2.4\times10^{-14}$ eV. Thus the matched recovery as shown in Fig.~\ref{fig:matched} applies to both boundary conditions.

\newpage
\textbf{Lattice mismatch: full finite spectra}
\begin{figure}[h]
    \centering
    \includegraphics[width=0.9\linewidth]{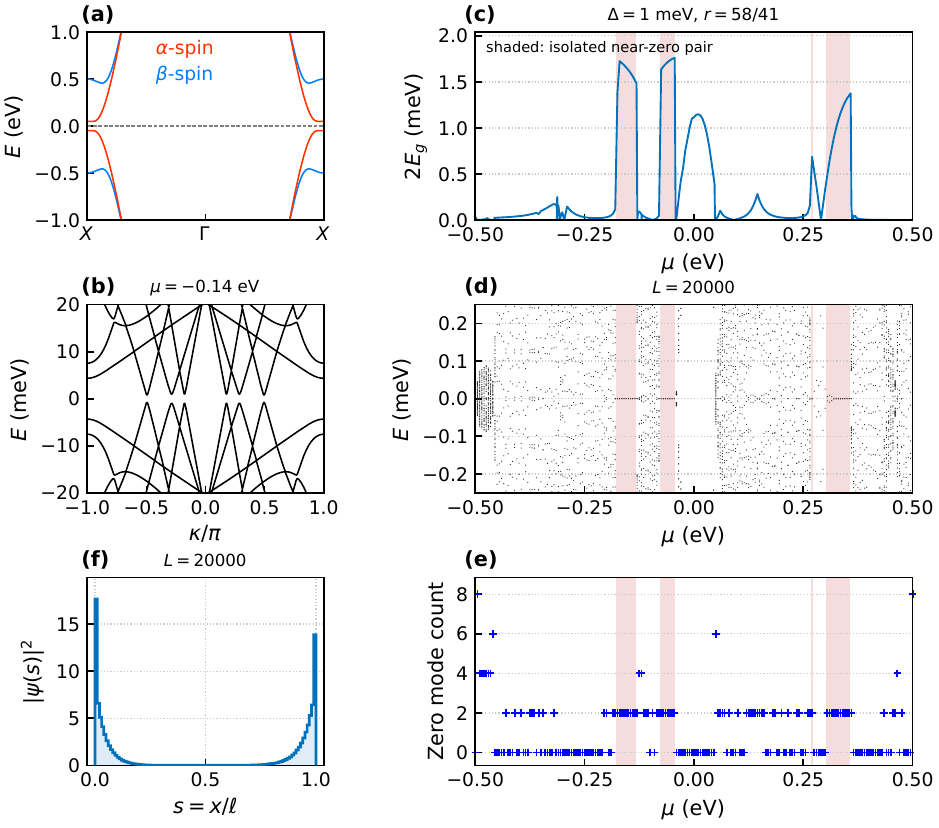}
    \caption{\label{fig:mismatched20k} Calculations accounting for lattice mismatch between ZGNR and Ising superconductor with the ratio $58/41\sim 3.48/2.46$ parameterizes the lattice mismatch. Rest of the parameters in the calculations are same as Fig. 2 of the main text. (a) Bandstructure of the ZGNR in transverse in-plane electric showing magnetic bands where $\alpha$-spin bands are split from $\beta$ spin bands. (b) BdG spectrum of the heterostructure obtained at $\mu = -0.14\ \text{eV}$. 
  (c) The real space distribution of the zero energy mode. 
  (d) Bogoliubov excitation gap for the heterostructure with periodic boundary condition as a function of chemical potential $\mu$ shows spectral gaps and topological phase transitions indicated by gap closings and openings.
  (e) Low-energy spectrum near zero energy for a finite-length system (L=20000 unit cells). The gapped continuum of bulk states accurately mirrors the PBC gap shown in (d). Crucially, two in-gap states emerge, pinned at zero energy in the topological phase. These are the Majorana zero modes (MZMs) localized at the system's ends.
  (f) Number of zero-energy states, determined by counting eigenvalues with an energy $|E| < 10^{-5}$ eV. The plot confirms that the system hosts either zero or exactly two zero-energy modes, consistent with the expected topological invariant for this phase.}
\end{figure}

\begin{figure}[h]
    \centering
    \includegraphics[width=0.9\linewidth]{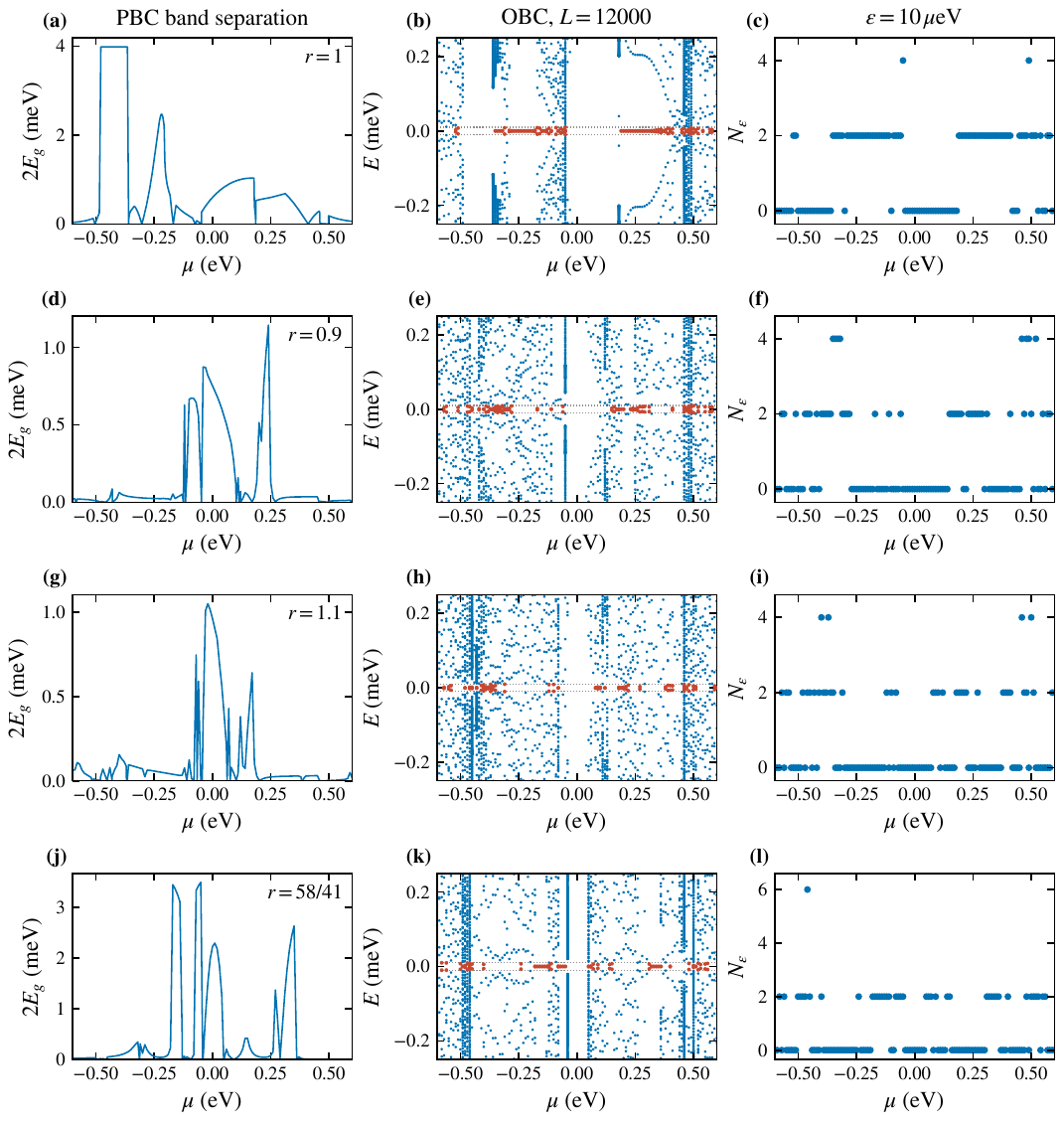}
    \caption{\label{fig:mismatched} Local-contact results at $\Delta=2$ meV and $L=12000$, for $r=1,0.9,1.1,58/41$. Left: same-model PBC band separation $2E_g$. Centre: every OBC eigenvalue in $|E|<0.25$ meV. Right: the complete $\varepsilon=10\,\mu$eV signed-state count. Red OBC points are inside this counting window. Each row uses 121 chemical potentials spanning $-0.60\le\mu\le0.60$ eV at $0.01$ eV spacing. Bulk calculations include this grid and directly evaluated endpoint gap closings; the matched bulk curve additionally uses its refined mesh. The finite spectral vertical scale is common to all rows; bulk vertical scales vary to resolve the different gaps.}
\end{figure}

The ratio $58/41=3.48/2.46$ uses nominal NbSe$_2$ and graphene lattice periods [3]. A longitudinal supercell contains 58 GNR and 41 SC periods, with common length $58a_g\simeq14.27$ nm. With the present $a_g=\sqrt3a_{\rm CC}$, the corresponding SC period is approximately $3.479$ \AA. Ratios $0.9$ and $1.1$ test both signs of a smaller mismatch. Both in-plane SC coordinates change, while the intralayer hopping amplitudes stay fixed, isolating geometric interface registry and site-density effects within the effective strip model.

\newpage
Now we also study various different lattice constant mismatches and their effects on topological regions and MZMs.
\textbf{Protection gap and finite-length evolution}
\begin{figure}[h]
    \centering
    \includegraphics[width=1.0\linewidth]{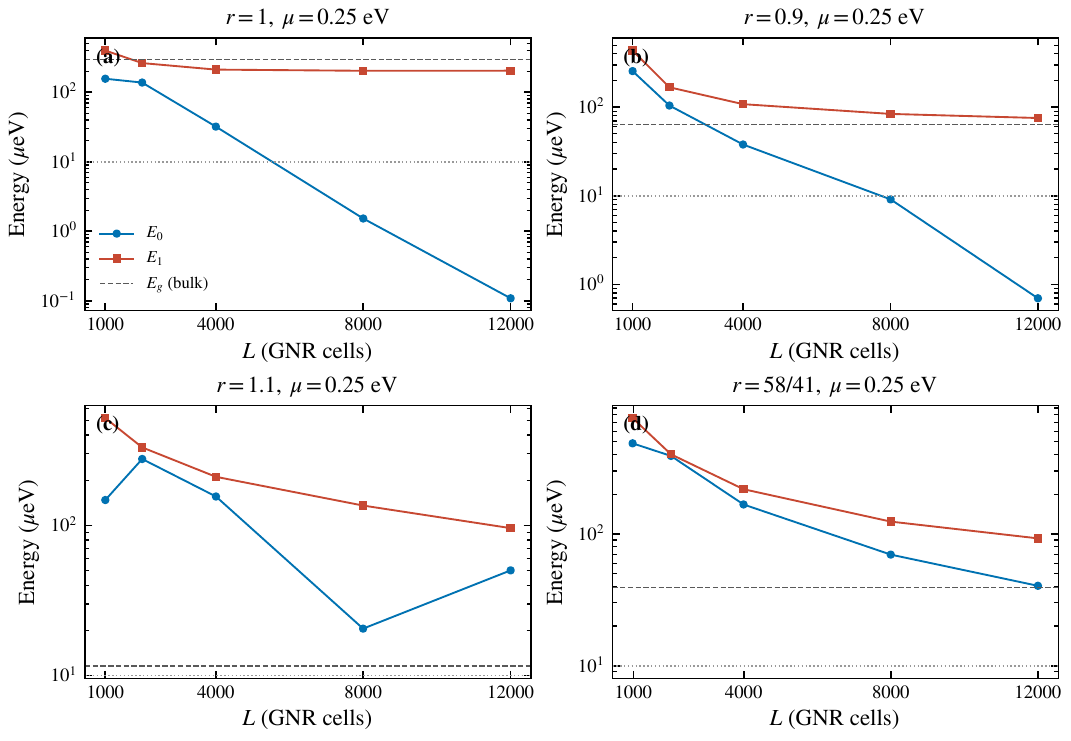}
    \caption{\label{fig:mismatched} Lowest two positive OBC energies versus length at the common chemical potential $\mu=0.25$ eV and $\Delta=2$ meV. Blue circles denote $E_0$ and red squares $E_1$. Dashed lines show each model's bulk $E_g$; dotted lines indicate $10\,\mu$eV. Lengths are chosen in the same supercell-termination class as $L=12000$. The GNR end Hartree profiles and interface parameters remain fixed.}
\end{figure}

 Lowest two positive OBC energies versus length at the common chemical potential $\mu=0.25$ eV and $\Delta=2$ meV. Blue circles denote $E_0$ and red squares $E_1$. Dashed lines show each model's bulk $E_g$; dotted lines indicate $10\,\mu$eV. Lengths are chosen in the same supercell-termination class as $L=12000$. The GNR end Hartree profiles and interface parameters remain fixed.

For $r=p/q$, the bulk calculation repeats a supercell of $p$ GNR and $q$ SC periods, with $E_g=\min_{\kappa,n}|E_n(\kappa)|$ and Bloch phase $\kappa=k_xpa_g$. At width three, the BdG dimension is $24(p+q)$: $48$, $456$, $504$ and $2376$ for the four ratios. The transverse direction remains a finite strip. The bulk operator uses the interior Hartree field and the same interface registration as the finite chain.

A longer chain reduces the envelope of end-state overlap, but it does not restore a bulk gap that the interface geometry has reduced. At this common chemical potential, the mismatch changes the protection gap as well as the finite-length splitting. In particular, the $r=1.1$ and $58/41$ examples have $E_0>E_g$ and $E_0>10\,\mu$eV at $L=12000$; these finite chains do not resolve an isolated near-zero pair at $\mu=0.25$ eV. A fixed-length threshold count alone therefore does not establish or exclude a bulk topological phase.

\begin{figure}[h]
    \centering
    \includegraphics[width=0.9\linewidth]{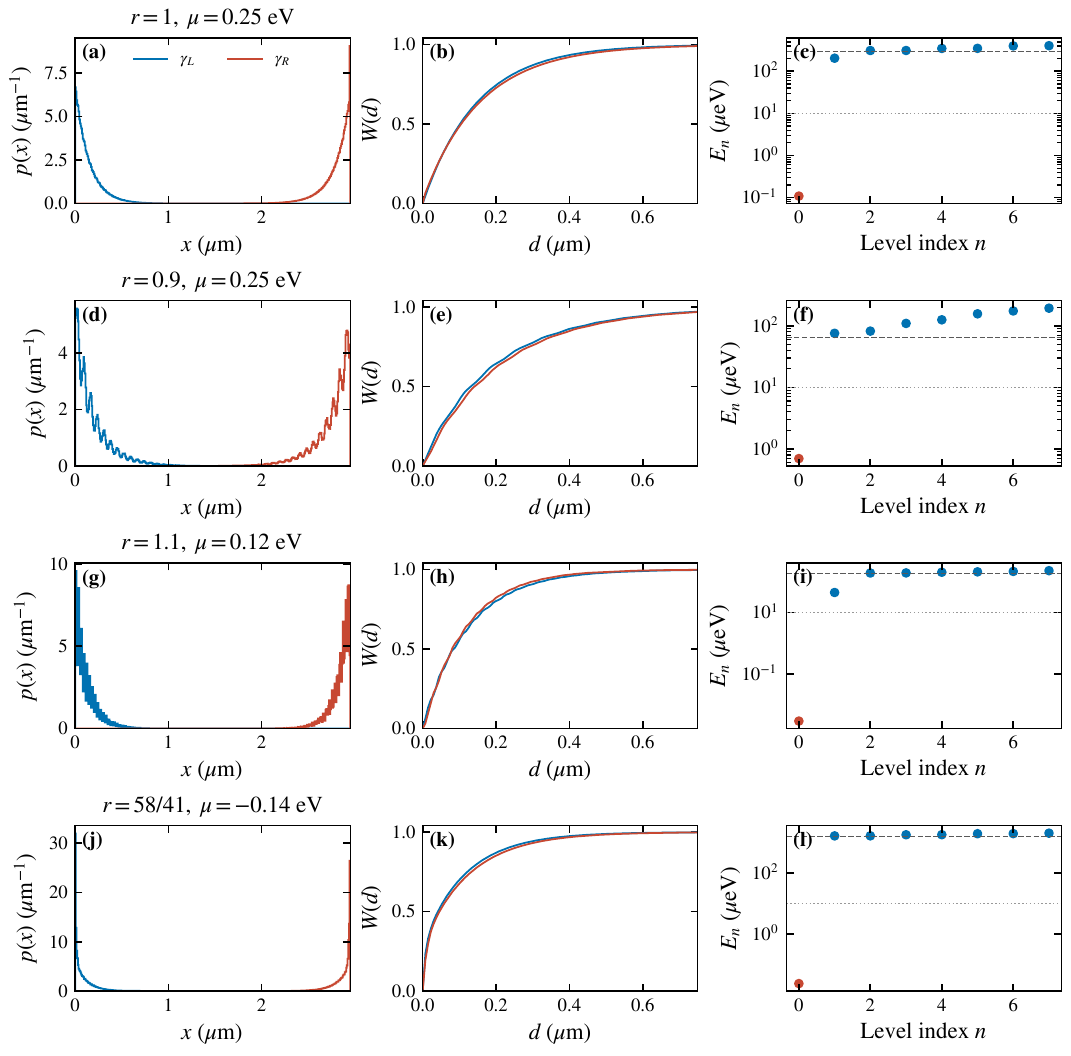}
    \caption{\label{fig:mismatchedwf} Representative long-chain wavefunctions at $\Delta=2$ meV, $L=12000$. Rows specify $(r,\mu)$ in the first-column titles. Left: spatial densities of the normalized self-conjugate components $\gamma_L$ (blue) and $\gamma_R$ (red). Centre: probability accumulated from the corresponding end over distance $d$. Right: the lowest eight positive OBC energies $E_0\le E_1\le\cdots\le E_7$, plotted against their level index $n$ on a logarithmic energy scale. Dashed lines denote $E_g$ and dotted lines the $10\,\mu$eV threshold. The negative particle-hole partners are implicit; one positive level below threshold corresponds here to two signed states in the finite-spectrum count.}
\end{figure}

\begin{figure}[h]
    \centering
    \includegraphics[width=0.9\linewidth]{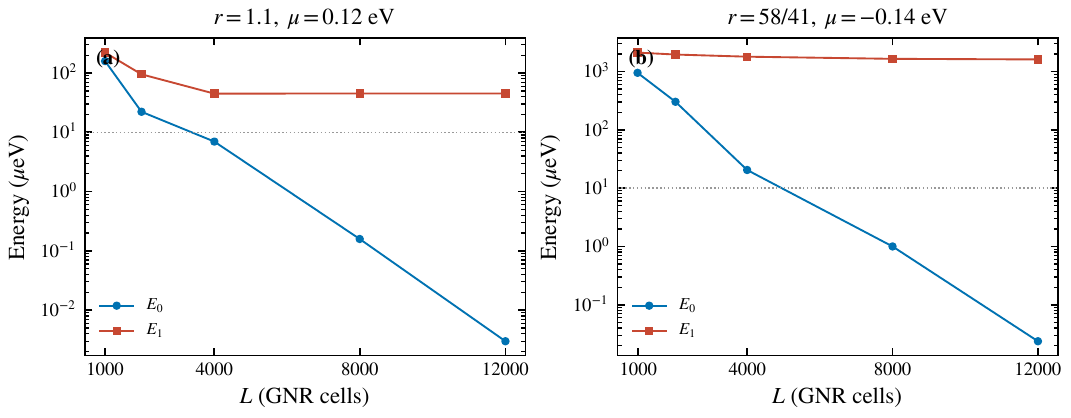}
    \caption{\label{fig:mismatchedwflength} Termination-controlled evolution of the two mismatched examples in the last two rows of Fig.~\ref{fig:mismatchedwf}. All model parameters and end Hartree profiles remain fixed as complete supercells are added. The lowest pair approaches zero and remains isolated from the next excitation. The dotted line is the fixed $10\,\mu$eV counting threshold.}
\end{figure}

The bulk gap is the onset of the infinite strip's positive bulk spectrum, not an upper bound on finite-chain energies. Discrete bulk-like excitations can lie above it, while boundary subgap levels can lie below it. We therefore use the actual $E_1$ when assessing energy isolation. Figures~\ref{fig:mismatchedwf},\ref{fig:mismatchedwflength} examine other chemical potentials where the mismatched system supports well-separated end components. Together, these results demonstrate surviving end-state regions after mismatch, rather than an absence of mismatch effects at every chemical potential.

\newpage
\textbf{End-state evidence in the local-contact mismatch model}

Probabilities sum over both materials, spins, orbitals and Nambu components. For noncoincident lattices we use physical coordinates: in each of 400 uniform position bins, $p(x)=P_{\rm bin}/\Delta x$, so $\int p(x)\,dx=1$. The cumulative quantity is $W(d)=\int_0^d p_{
\rm end}(s)\,ds$. The components are constructed from the lowest particle-hole subspace $\pm E_0$ as in Fig.~\ref{fig:Endstates}; $E_1$ is the next excitation, not the energy of the other end component. Self-conjugacy by itself follows from this construction; the distinct end peaks, isolation in energy, open bulk gap and length dependence together supply the physical evidence.

The last two rows use $\mu=0.12$ and $-0.14$ eV, respectively, rather than the common $\mu=0.25$ eV of Fig.~\ref{fig:mismatched}. They establish that suitable end-state regions survive mismatch: in each example $E_0\ll E_1$, the two components peak at opposite ends, and the bulk remains gapped. These selected points do not imply unchanged phase boundaries or protection gaps over the full chemical-potential range.

\textbf{Length controls for the mismatched end-state regimes}

\textbf{Interpretation.} The local-contact extension recovers the matched main-text model exactly and supports well-separated end-state pairs in representative regions for both signs of mismatch, including the $58/41$ ratio. Mismatch nevertheless changes the gap and the chemical-potential locations at which a finite chain clearly resolves such states. Additional threshold states and poorly localized low-energy states are not identified as an isolated Majorana pair on the basis of that count alone.

\textbf{Numerical control.} The finite calculations use sparse real-basis representations of the unchanged BdG operator. Shift-invert spectral slices provide all states in each displayed interval; original physical-basis residuals are below $10^{-10}$ eV. Independent symmetric block-congruence counts, using opposite elimination directions and perturbed interval boundaries, certify the displayed spectral multiplicities. The $10\,\mu$eV counts are checked separately. Small-system complete diagonalizations and independent long-chain solvers validate the construction and energies. Bulk minima are refined in every sampled local basin with successive momentum meshes and independent complete-BdG low-energy checks. The three-row strip, imposed pairing and frozen GNR Hartree background define the effective model's scope. A material-specific prediction requires a calibrated interface and substrate-width analysis.

\newpage
\subsection{Supplementary Note 8: Spatially modulated tunnel coupling between ZGNR and SC}
In the main text, we have assumed lattice matching between ZGNR and the superconductor for simplicity. 
At, the minimal level, the lattice constant mismatch can be modeled as spatially varying interlayer tunneling between the ZGNR and superconductor. To qualitatively account for this effect, we performed a finite-size calculation in which the tunneling coupling between the ZGNR and the Ising superconductor is modulated along the periodic direction of the ZGNR. Specifically, we take the modulated tunneling
\begin{equation}
    w' = w \cos\left(\frac{2\pi x}{ba} \right),
\end{equation}
with $w$ the local single-particle tunneling term as in Eq.7 in the main text. $a$ is the lattice constant of the ZGNR in the periodic direction, and the integer $b$ represents the period of this modulation in terms of the number of repeated unit cells before the tunneling amplitude returns to its original value. This simple real-space modulation effectively mimics the local variation of the tunneling strength caused by the lattice constant mismatch between the layers.

\begin{figure}[h]
    \centering
    \includegraphics[width=0.5\linewidth]{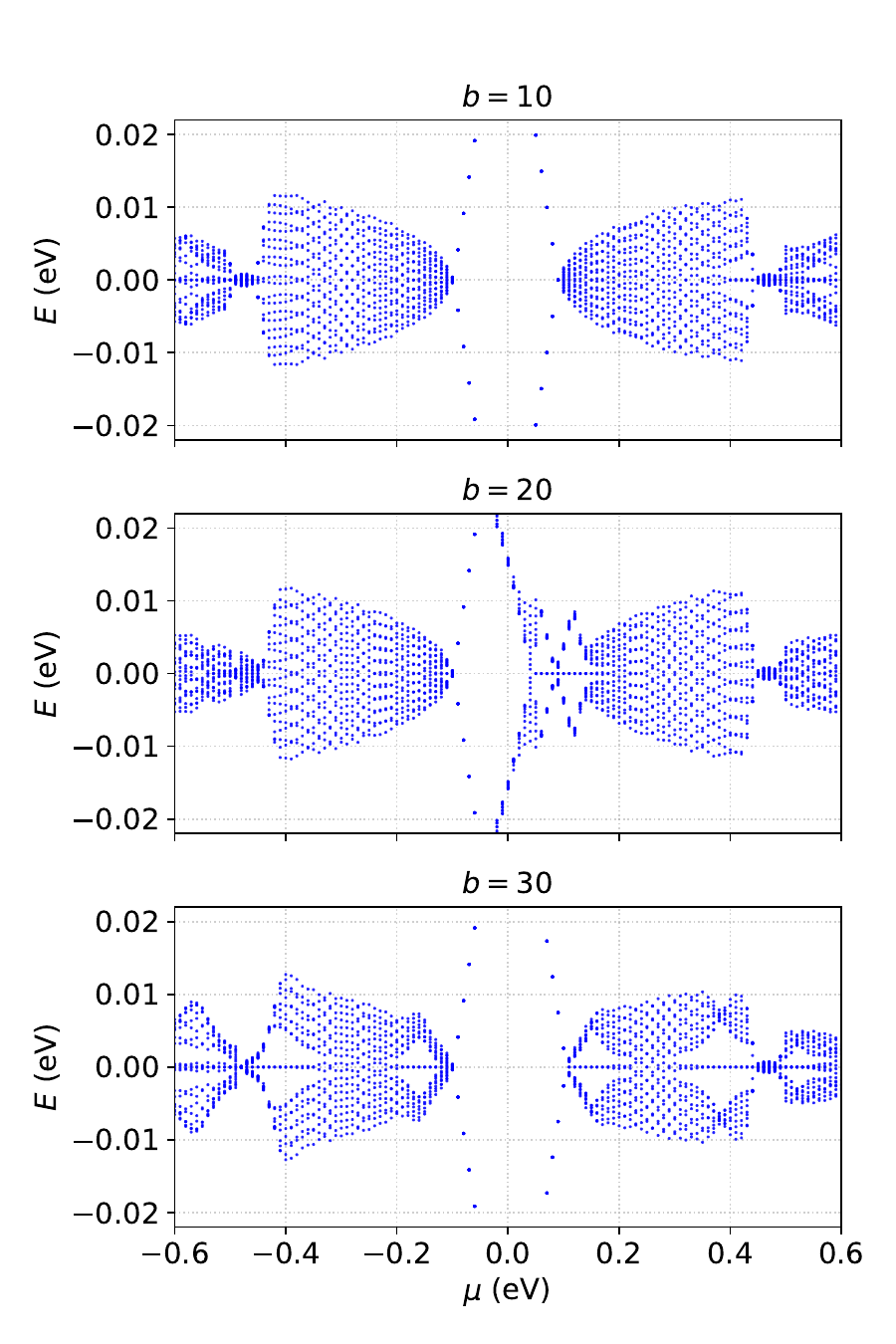}
    \caption{Finite-sized energy spectrum for proximity coupled ZGNR with $W=3, L=1000, U=t$ and $F=0.1$ V/Å, for spatial modulation of interlayer tunneling of different periods.}
    \label{fig:sm_modulation}
\end{figure}

In Fig~\ref{fig:sm_modulation}, we present the energy spectrum of the finite system with different modulation period $b$, and the MZMs are clearer for systems with smoother modulations where the topological phases are retained. While for strong and rapid modulation, it could disrupt the bands and thus destabilize the MZMs.


\newpage
\subsection{Supplementary Note 9: Half-metallicity and MZMs for $U < t$}
In the main text, we have shown that the ZGNR under an external field exhibits a spin-polarized half-metallic state at the Fermi level, originating from strong on-site electron-electron interaction $U$, such that $U/t \approx 1$. Here, we show the robustness of this phase when the interaction $U$ is reduced due to screening effects that might be present from coupling to the superconductor. To address this, we have performed additional mean-field calculations to determine the stability of the half-metallic phase and, correspondingly, the existence of Majorana zero modes.

\begin{figure}[ht]
    \centering
    \begin{minipage}[b]{0.48\textwidth}
        \centering
        \begin{picture}(0,0)
        \end{picture}
        \includegraphics[width=\textwidth]{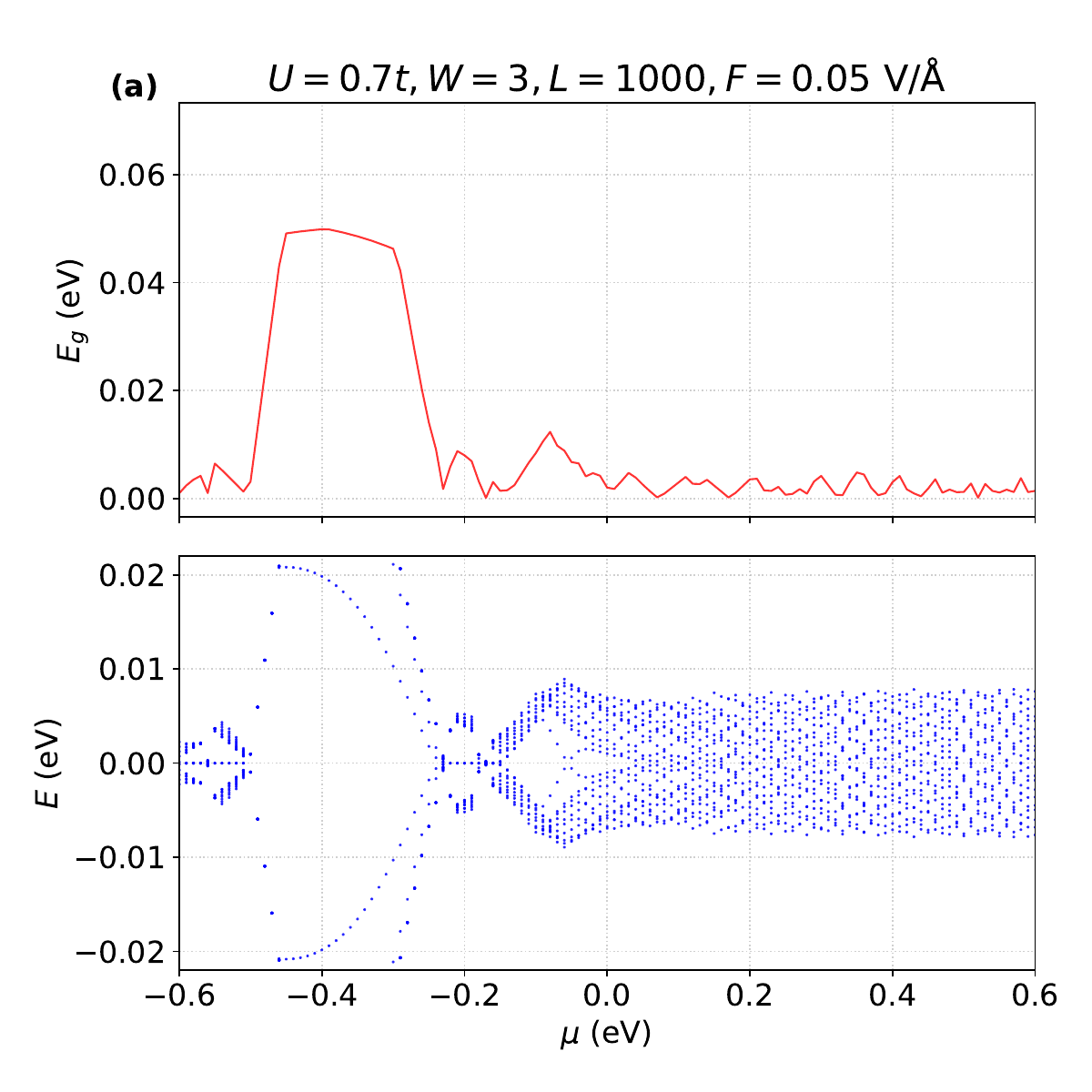}
    \end{minipage}
    \hfill
    \begin{minipage}[b]{0.48\textwidth}
        \centering
        \begin{picture}(0,0)
        \end{picture}
        \includegraphics[width=\textwidth]{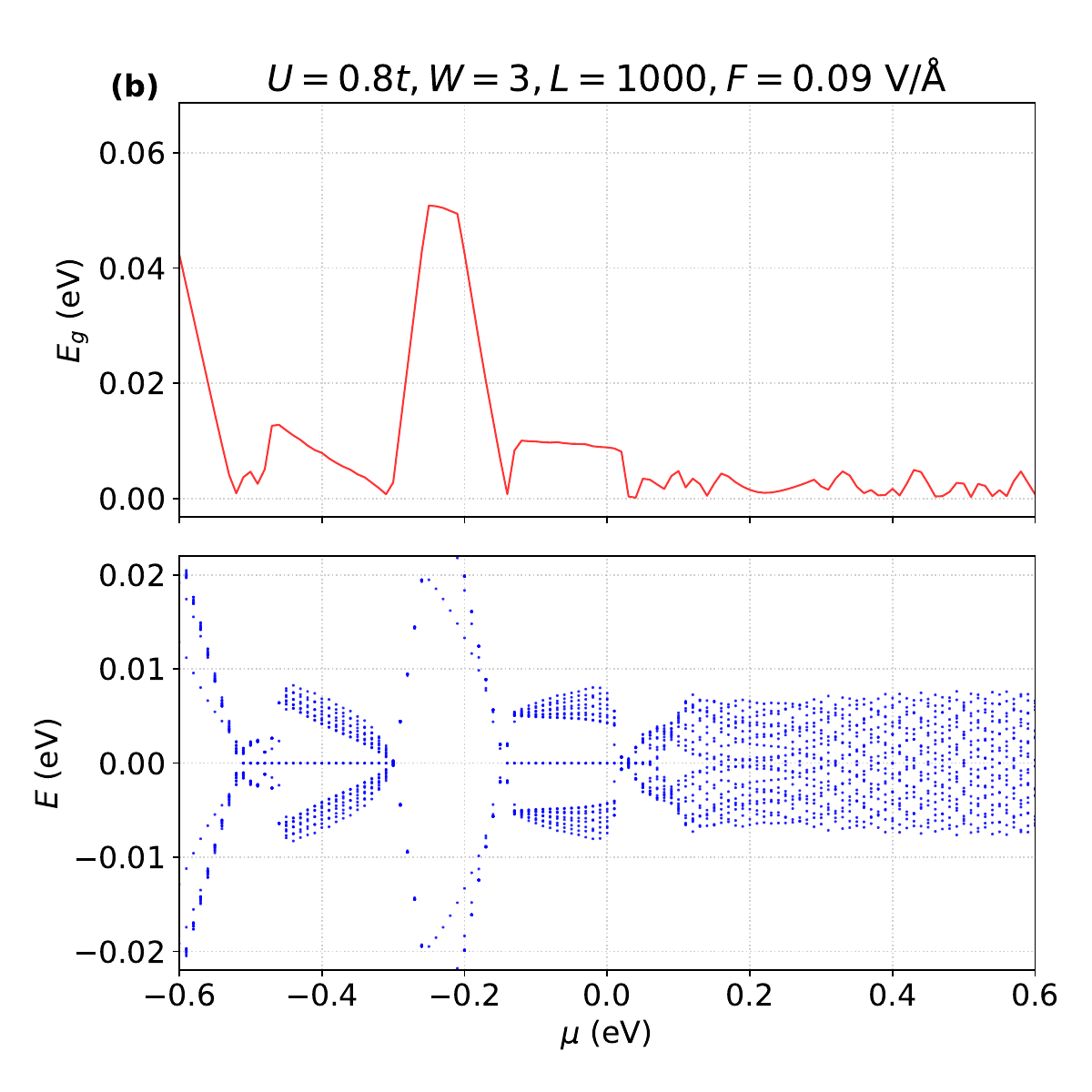}
    \end{minipage}
    \caption{(a) Bogoliubov excitation gap $E_g$ for the heterostructure with periodic boundary condition and low-energy spectrum (20 states closest to zero energy) for a finite-length system (L=1000 unit cells) for $W=3$ ZGNR with $U/t=0.7$ and $F=0.05$ V/Å. (b) Same as (a) but at for $U/t=0.8$ and $F=0.09$ V/Å.}
    \label{fig:sm_finite_uot}
\end{figure}

\begin{figure}[ht]
    \centering
    \includegraphics[width=0.9\linewidth]{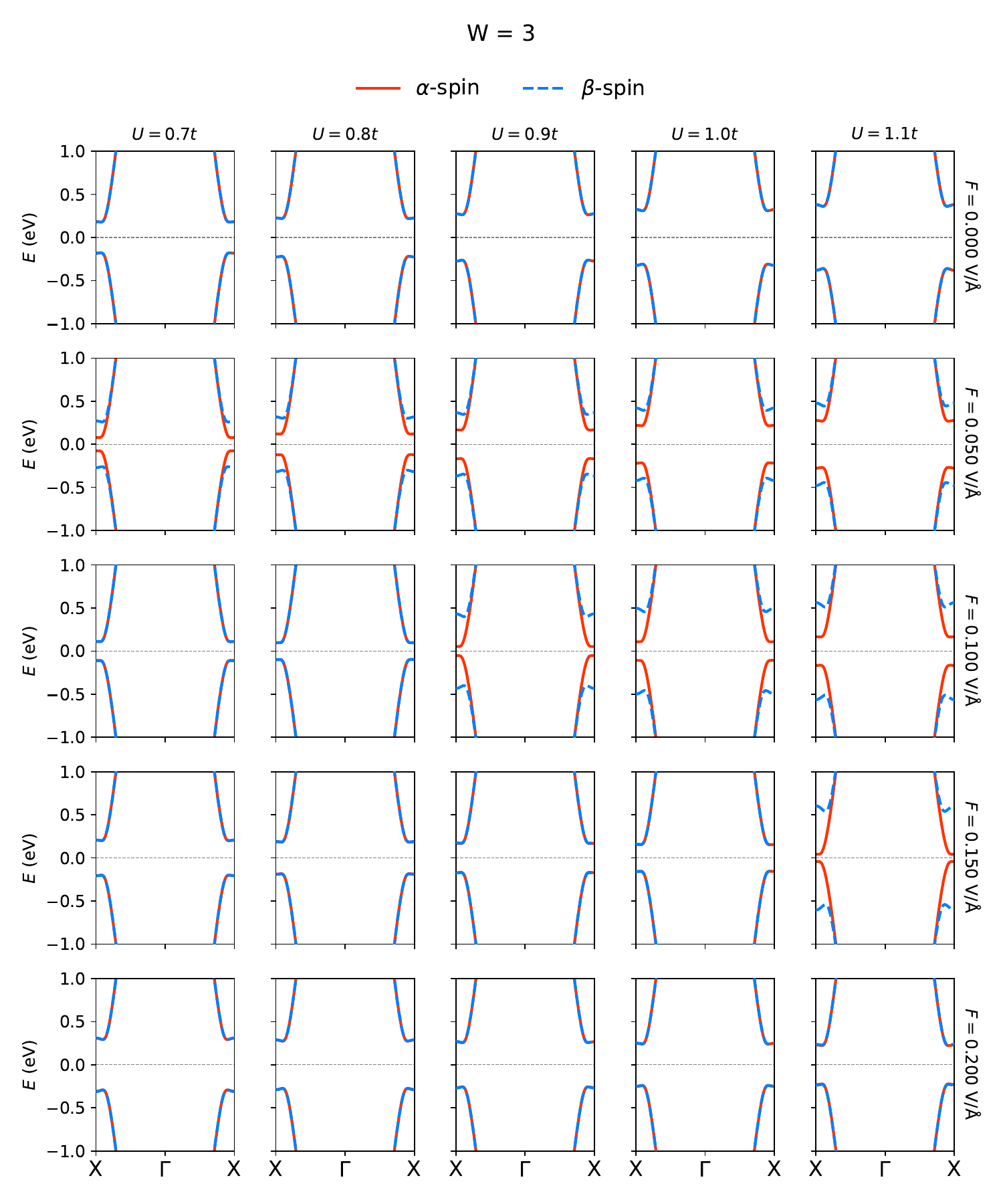}
    \caption{Spin-polarized electronic band structures of ZGNR for $W=3$, with different external fields and $U/t$ ratios.}
    \label{fig:sm_ZGNR_W3}
\end{figure}

\begin{figure}[ht]
    \centering
    \includegraphics[width=0.9\linewidth]{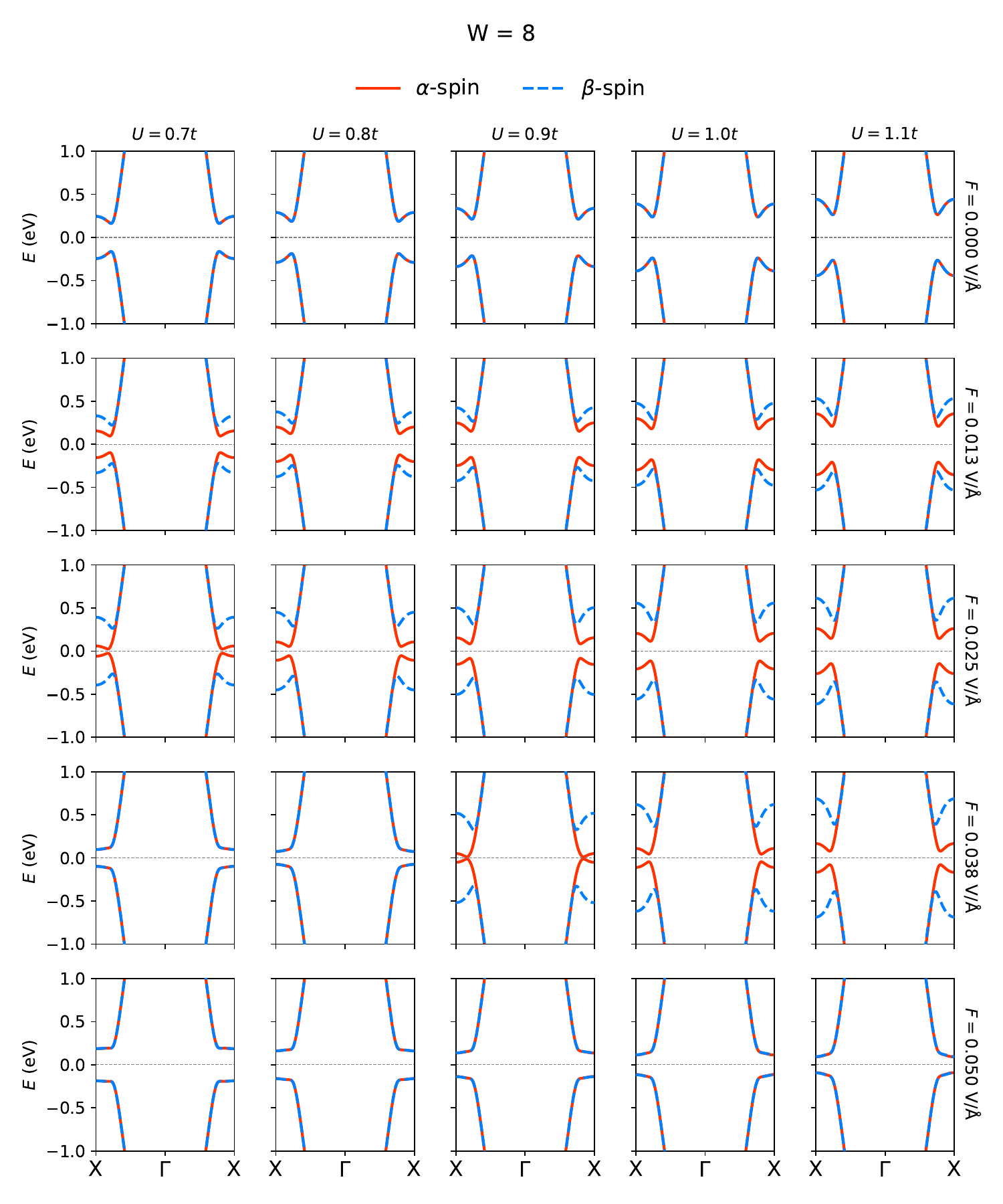}
    \caption{Spin-polarized electronic band structures of ZGNR for $W=8$, with different external fields and $U/t$ ratios.}
    \label{fig:sm_ZGNR_W8}
\end{figure}

\begin{figure}[ht]
    \centering
    \includegraphics[width=0.9\linewidth]{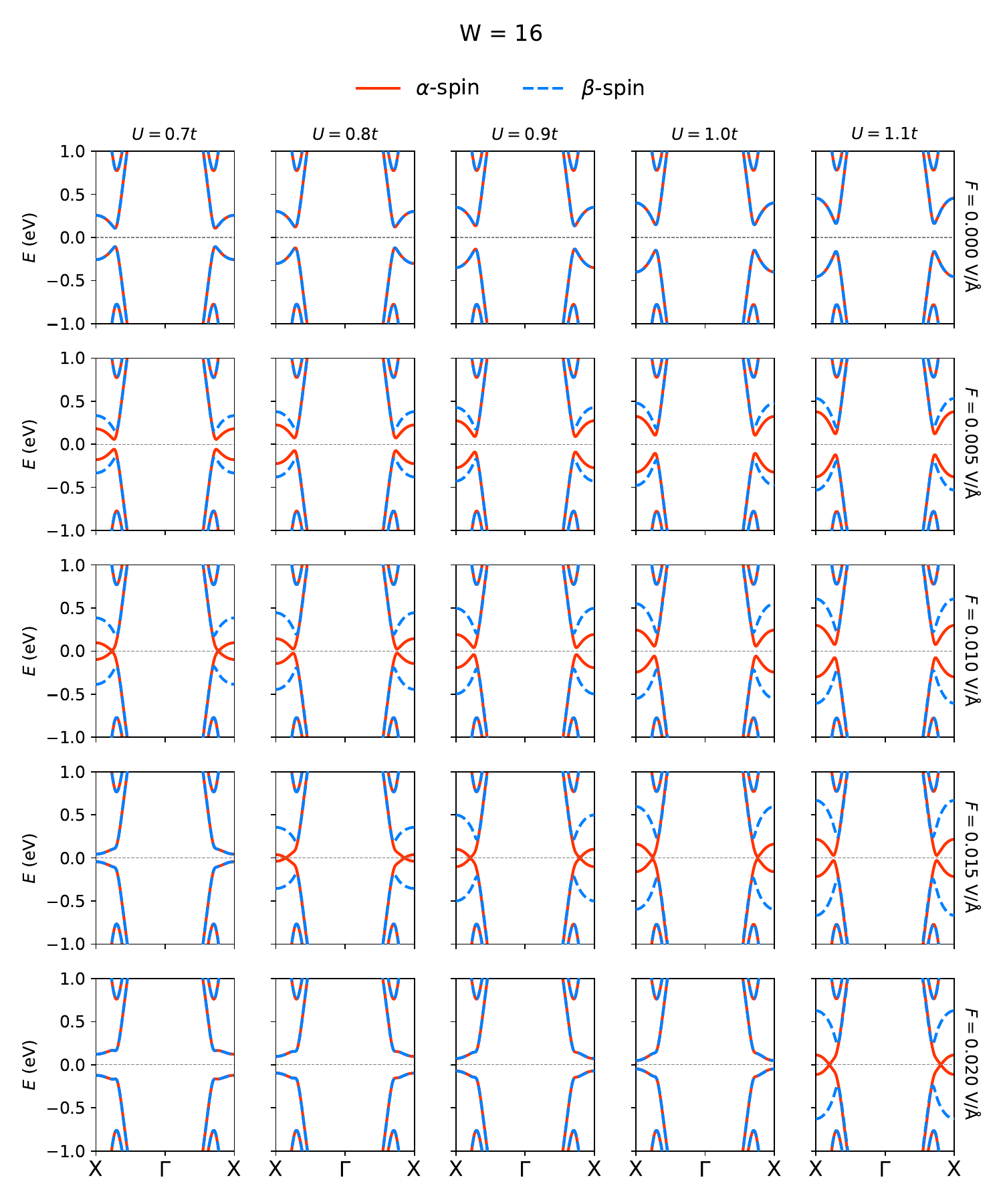}
    \caption{Spin-polarized electronic band structures of ZGNR for $W=16$, with different external fields and $U/t$ ratios.}
    \label{fig:sm_ZGNR_W16}
\end{figure}

From Fig.~\ref{fig:sm_ZGNR_W3} to Fig.~\ref{fig:sm_ZGNR_W16}, we show the spin-polarized electronic band structure of ZGNR with different widths $W$ in different transverse in-plane electric fields $F$, for different $U/t$ ratios. While the half-metallic phase has a complex dependence on these parameters, we can see that for each of the width we present, we can find a suitable electric field to achieve the half-metallicity down to $U=0.7t$. In particular, we also select representative fields for $U=0.7t$ and $U=0.8t$ to perform the finite-sized calculations for ZGNR with $W=3$, as shown in Fig.~\ref{fig:sm_finite_uot}. The PBC bandgap and the finite spectrum match with significant regions of MZMs, when the ZGNR is in the half-metallic regions.

\newpage

\newpage

\newpage

\bibliography{GNR_Majorana_bibliography}